\documentclass[10pt,journal,compsoc]{IEEEtran}

\usepackage[algoruled,vlined,linesnumbered]{algorithm2e}
\usepackage{graphicx}
\usepackage{array}
\newcolumntype{C}[1]{>{\centering}p{#1}}
\usepackage{amsmath}
\usepackage{amssymb}
\usepackage{booktabs}
\usepackage{enumerate}
\usepackage{arydshln}
\usepackage{bm}
\usepackage{cases}
\usepackage{color}
\usepackage{xcolor}
\usepackage{comment}
\usepackage{dsfont}
\usepackage{bbding}
\usepackage{subfigure}

\usepackage{wrapfig}
\definecolor{mycolor}{RGB}{206,255,255}    
\definecolor{mycolor1}{RGB}{242,232,227}     

\definecolor{mycolor-MLMC-blue}{RGB}{255,255,255}    
\definecolor{mycolor-MLMC-yellow}{RGB}{255,255,255}     

\usepackage[colorlinks,linkcolor=pink]{hyperref}

\ifCLASSOPTIONcompsoc
  \usepackage[nocompress]{cite}
\else
  \usepackage{cite}
\fi

\ifCLASSINFOpdf

\else

\fi

\begin{document}

\title{Blind Super-Resolution via Meta-learning and Markov Chain Monte Carlo Simulation}

\author{Jingyuan~Xia,~\IEEEmembership{}
        Zhixiong~Yang,~\IEEEmembership{}
        Shengxi~Li,~\IEEEmembership{Member,~IEEE},
        Shuanghui~Zhang,~\IEEEmembership{Member,~IEEE},
        Yaowen~Fu,~\IEEEmembership{}
        Deniz~Gündüz,~\IEEEmembership{Fellow,~IEEE}
        and~Xiang~Li,~\IEEEmembership{}
        
\IEEEcompsocitemizethanks{
\IEEEcompsocthanksitem Jingyuan~Xia, Zhixiong~Yang, Shuanghui~Zhang, Yaowen~Fu and Xiang~Li are with College of Electronic Engineering, National University of Defense Technology, Changsha, 410073, China. \protect\  Shengxi Li is with College of Electronic Engineering, Beihang University, Beijing, 100191, China. \protect\  Deniz Gündüz is with the Department of Electrical and Electronic Engineering, Imperial College London, SW7 2AZ, UK, and the `Enzo Ferrari' Department of Engineering, University of Modena and Reggio Emilia, Italy. \protect
\IEEEcompsocthanksitem E-mail: (j.xia16, shengxi.li17, d.gunduz)@imperial.ac.uk,\\
(yzx21, zhangshuanghui, fuyaowen, lixiang)@nudt.edu.cn. \protect\
\IEEEcompsocthanksitem Jingyuan~Xia and Zhixiong Yang contributed equally to this work (Corresponding authors: Zhixiong Yang and Shuanghui Zhang). \protect\
\IEEEcompsocthanksitem This paper has been accepted for publication in IEEE Transactions on Pattern Analysis and Machine Intelligence. \protect\
}
}

\IEEEtitleabstractindextext{%
\begin{abstract}
Learning-based approaches have witnessed great successes in blind single image super-resolution (SISR) tasks, however, handcrafted kernel priors and learning based kernel priors are typically required. In this paper, we propose a Meta-learning and Markov Chain Monte Carlo (MCMC) based SISR approach to learn kernel priors from organized randomness.
In concrete, a lightweight network is adopted as kernel generator, and is optimized via learning from the MCMC simulation on random Gaussian distributions. This procedure provides an approximation for the rational blur kernel, and introduces a network-level Langevin dynamics into SISR optimization processes, which contributes to preventing bad local optimal solutions for kernel estimation.
Meanwhile, a meta-learning-based alternating optimization procedure is proposed to optimize the kernel generator and image restorer, respectively. 
In contrast to the conventional alternating minimization strategy, a meta-learning-based framework is applied to learn an adaptive optimization strategy, which is less-greedy and results in better convergence performance. These two procedures are iteratively processed in a plug-and-play fashion, for the first time, realizing a learning-based but plug-and-play blind SISR solution in unsupervised inference. 
Extensive simulations demonstrate the superior performance and generalization ability of the proposed approach when comparing with state-of-the-arts on synthesis and real-world datasets. The code is available at \href{https://github.com/XYLGroup/MLMC}{https://github.com/XYLGroup/MLMC}.
\end{abstract}

\begin{IEEEkeywords}
Blind single image super-resolution, Markov Chain Monte Carlo simulation, Meta-learning.
\end{IEEEkeywords}}

\maketitle

\IEEEdisplaynontitleabstractindextext

\IEEEpeerreviewmaketitle

\section{Introduction}\label{sec:introduction}

\IEEEPARstart{S}{ingle} image super-resolution (SISR) plays a crucial role in image processing society. It tends to reconstruct high-resolution (HR) images from the low-resolution (LR) observations. With the fact that the degradation model is typically unknown in real-world scenarios, growing studies begin to predict the blur kernels and the HR images simultaneously, known as the blind SISR problem. A common mechanism to solve the blind SISR problem is underlying an alternating optimization between two sub-problems, \emph{kernel estimation} and \emph{image restoration}, which are iteratively minimized until the HR image is restored. Kernel estimation is a pivotal step in solving blind SISR problems, which determines the HR image reconstruction performance, and thereby becomes the centrality study of this paper.

Most of the existing blind SISR approaches can be categorized as model-based approaches and learning-based approaches. Model-based approaches \cite{perrone2015clearer, ulyanov2018deep, bell2019blind, ren2020neural, emad2021dualsr, Yue2022blind} enjoy better generalization-ability because of the explicit degradation modeling and gradient-based solution scheme. However, these methods suffer from the ill-posedness and non-convexity of the blind SISR problem, and typically stuck at the bad local mode during optimization. Therefore, learning-based methods \cite{lai2018fast, anwar2020densely, liu2020residual, mei2020image, huang2020unfolding, wang2021unsupervised, liang2021flow, luo2022learning, kong2022reflash, saharia2022image, fang2023self} obtain popularity in recent years, which realize significantly better performance via substituting explicit degradation model by network-based learning procedure on a large amount of labeled data, therefore, providing data-driven image and kernel priors for solving the non-convex blind SISR optimization problem.  
Nevertheless, these methods are still restricted by high data dependency on training samples and dedicated time-consumption for model-training in real-world applications. 

In general, we claim that there is a trade-off between the reconstruction performance on individual tasks and the generalization capacity on different scenarios in blind SISR. 
On the one hand, stronger priors definitely bring better performance, as the learning-based methods achieve. Meanwhile, the generalized flexibility to real-world applications will be declined since strong priors typically lead to serious overfitting. On the other hand, rational kernel priors and an adaptive optimization strategy, instead of exhaustively minimizing each individual sub-problem, are necessary to handle the intrinsic non-convexity and ill-posedness of blind SISR problems.
The latest meta-learning-based optimization algorithms \cite{li2020continuous, chen2021multiagent, xia2021meta, xia2022metalearning} have substantiated significant advances on solving non-convex optimization problems, in particular, for those with an alternating minimization (AM) framework, just as blind SISR problem does. 
In light of the meta-learning mechanism that learns to extract mutual knowledge for a ``bird's eye view" on global scope optimization, these algorithms achieve considerably better convergence performance on non-convex geometry. 
The common idea of meta-learning-based optimization strategies lies on incorporating organized randomness to prevent trapping into bad local optima, and a network-based optimizer is meta-learned across iterations to ensure the incorporated random disturbance following the primary objective of the optimization problem. In this instance, this paper strives to realize a novel blind SISR solution scheme, in which rational kernel priors and adaptive optimization strategy are gained through learning from organized randomness without cumbersome training in advances as well as data dependency. 

In this paper, we propose a Meta-learning and Markov Chain Monte Carlo (MLMC) based approach to solve the blind SISR problem.
It establishes a two-phase SISR optimization, including a \emph{Markov Chain Monte Carlo kernel approximation} (MCKA) phase and a \emph{meta-learning based alternating optimization} (MLAO) phase.
In the MCKA phase, CMC simulation on random Gaussian distributions is proposed to substitute traditional model-based or learning-based kernel priors, which are either handcrafted or deterministic via pre-training. 
In the contrary, the proposed MCKA aims to learn task-general kernel priors from random Gaussian distributions, and thereby achieves pre-training and labeled data free. A Markov Chains system modeling on optimization trajectory of the blind SISR problem is proposed to provide \emph{organized randomness}, which allows the Monte Carlo simulations to sample Gaussian distributions with a global scope on LR image reconstruction errors, therefore, providing rational kernel priors for better convergence performance.  
In concrete, a lightweight network is iteratively optimized in an unsupervised manner using MCMC simulations to approximately generate a blur kernel, which is trained across random Gaussian distributions. In this way, good generalization properties and flexibility towards arbitrary SISR tasks are ensured, and this organized randomness leads to a relaxation that prevents the kernel estimation from trapping into bad local modes.
Moreover, the MLAO phase refines the blur kernel and restores the HR image, alternatively, with respect to the task-specific LR observation. Instead of the exhaustive minimization on each individual sub-problem, the optimizer for kernel estimation is meta-learned by minimizing the accumulated LR image reconstruction errors over MLAO iterations. 
Then, an adaptive and effective optimization strategy for better convergence performance is obtained by learning the mutual knowledge of solving a sequence of MLAO sub-problems. 
These two phases are operated alternatively to realize a balance between task-general relaxation via learning from organized randomness and task-specific refinement.

\textcolor{black}{
The advantages of the proposed MLMC are listed below: 
i) \emph{Plug-and-play}. The MCKA phase effectively substitutes the deterministic kernel priors by a \textcolor{black}{plug-and-play} learning process through MCMC simulation on  Gaussian distributions. This results in negligible time-consumption and circumvents the cumbersome demand of \textcolor{black}{labeled training samples}. 
ii) \emph{Better convergence performance}. \textcolor{black}{Though without training in advance,} the MLAO phase achieves an adaptive non-convex optimization strategy that converges to better optimum for the estimated blur kernels. Meanwhile, the MCKA phase can be regarded as a network-level Langevin dynamics \cite{welling2011bayesian, arvinte2022mimo, Yue2022blind} towards the kernel estimation in the MLAO phase, which provides a rational and trainable random disturbance with the kernel estimation via learning from random Gaussian distributions.
This also ensures the convergence of kernel estimation.
iii) \emph{Stronger generalization capacity and flexibility}. The obtained kernel priors from the MCMC simulation are loose and dynamic, which endows better generalization capability on different degradation scenarios. Besides, the \textcolor{black}{plug-and-play} fashion evades the data dependency and re-training requests, therefore, significantly improving the flexibility in practical applications.
The main contributions of this paper are listed below:
\begin{itemize}
\item A universal statistic framework for the network-based degradation model is proposed to elaborate the network-based approach for the SISR problem. On the basis of this, a MCMC simulation model on random Gaussian distributions with a glimpse on LR image reconstruction error is established to elaborate a new kernel approximation phase.
\item  Different to the commonly-applied pre-training or manually designed kernel priors, a random kernel learning scheme replaces the ordinary kernel priors, which realizes a \textcolor{black}{learning-based but plug-and-play kernel prior generation paradigm and contributes to a commonly network-level Langevin dynamics optimization for convergence improvements.}
\item A meta-learning-based adaptive strategy is constructed to solve the blind SISR problem. It learns to optimize the non-convex and ill-posed blind SISR problem in a less-greedy optimization strategy, and thereof ensures better convergence performance towards ground truth when only depending on the observed LR image. 
\item To the best of our knowledge, \textcolor{black}{MLMC is the first meta-learning-based plug-and-play SISR approach that achieves superior performance and can be directly applied to common kernel estimation tasks including isotropic and anisotropic Gaussian, non-Gaussian and motion kernels}, with competitive number of parameters, runtime and memory usage comparing to the state-of-the-art, as well as robustness to the noise.
\end{itemize}
}

\section{Related Work}\label{sec:related work}

\subsection{Blind SISR methods}\label{sec:Blind SISR methods}
The existing blind SISR approaches can be roughly categorised into model-based and learning-based approaches. 

\noindent
\textbf{Model-based methods.}
Most of the early model-based blind SISR approaches \cite{russell2003exploiting, glasner2009super, michaeli2013nonparametric, perrone2015clearer} merely aim to construct specific HR images priors with explicit formulations, such as gradient profile \cite{sun2008image}, hyper-Laplacian \cite{krishnan2009fast}, sparsity \cite{kim2010single} and total variation (TV) \cite{rudin1992nonlinear}, for better reconstruction performance. More recent studies begin to focus on kernel priors designing. For example, Jin \textit{et al}. \cite{jin2018normalized} propose a popular heuristic normalization as kernel prior which realizes better convergence. Yue \textit{et al}. \textcolor{black}{\cite{Yue2022blind} employ an explicit pre-defined Gaussian kernel model, achieving good robustness on noise scenarios. It iteratively optimizes the Gaussian kernel and the input noise via a gradient-based algorithm, which causes limitations towards capturing varying kernel categories.} In sum, model-based approaches are typically with good generalization-ability due to the explicit modeling on task-specific priors. Meanwhile, a significant performance decline will appear when the manually designed kernel priors slightly deviate from the degradation model in the application. Besides, due to the high ill-posedness and non-convexity of the blind SISR problem, model-based methods can easily get stuck at local optimums via the hand-crafted and rule-fixed gradient descent-based optimization algorithms. These urge researchers turn their attentions to learning-based alternatives for better convergence performance and more flexibility. 

\noindent
\textbf{Learning-based methods.}
The arise of deep learning has motivated the learning-based blind SISR approaches to learn kernel/image priors via network-based behavior. These models are typically pre-trained on paired LR/HR image samples to obtain image and kernel priors. Typically, end-to-end deep networks are adopted to formulate the degradation model \cite{dong2014learning}, and therewith bring about plenty of deep convolutional neural network (CNN) based SISR optimizers \cite{kim2016accurate, zhang2018learning, liu2020residual, xu2020unified, hui2021learning, kim2021koalanet, zhang2021designing, fang2022uncertainty, xia2022knowledge, zheng2022unfolded}. \textcolor{black}{Specifically, Liang \textit{et al}. \cite{liang2021mutual} propose an end-to-end deep network that estimates kernel for different patches in LR images via residual blocks. Fang \textit{et al}. \cite{fang2022uncertainty} establish a two-stage framework for solving blind SR tasks with statistical modeling on LR image, which learns mean and variance from LR image estimate kernel. To improve the generalization-ability towards diverse degradations of real images, Zhang \textit{et al}. \cite{zhang2021designing} design a practical degradation model with shuffled blur for the synthetic training data. More recently, Xia \textit{et al}. \cite{xia2022knowledge} propose a knowledge distillation-based implicit degradation estimator for blind SR tasks, allowing better generalization towards different degradations.}
Deep image prior (DIP) \cite{ulyanov2018deep} is one of the most well-known model that is capacity of learning image features in an unsupervised way with superior performance on solving computer vision tasks. However, it suffers from limited generalization-ability on different degradation models when dealting with blind SISR problems. 
In this instance, recent works \cite{hu2019meta, soh2020meta, li2021different, yu2021lite, xia2022metaBSR} introduce meta-learning for better generalization-ability via training across different samples in terms of image sizes, degradation categories and resolution ratios. Nevertheless, these methods demand a large amount of training samples and a time-consuming dedicated pre-training phase. Meanwhile, other works \cite{cornillere2019blind, zhang2019deep, zhang2020deep, huang2020unfolding, li2020efficient, liang2021mutual, fu2022kxnet} strive to expand the conventional model-based iterative solutions with learning-based behavior via unfolded deep learning algorithms. Gu \textit{et al}. \cite{gu2019blind} propose a network-based alternative optimization framework that simultaneously optimizes the blur kernel and HR image via deep models.   \textcolor{black}{Zhang \textit{et al}. \cite{zhang2020deep} formulate a fundamental deep unfolding framework for the SR task, which unfolds the MAP inference via a half-quadratic splitting algorithm. More recently, Liang \textit{et al}.\cite{liang2021flow} establish a flow-based kernel prior (FKP) model that realizes an improved performance through pre-training a kernel estimator as non-parametric priors, and incorporates it with the existing SISR approaches. Approaches in \cite{zheng2022unfolded, fu2022kxnet} formulate a deep-unfolding-based model on the basis of the half quadratic splitting algorithm and proximal gradient descent algorithm.}

In sum, the latest blind SISR methods illustrate a trend of reducing the dependency on training resources (pre-training stage, external training data) and strengthening the generalization-ability on varying degradation categories. Although the exsiting works, such as FKP-DIP \cite{liang2021flow}, in which the necessity of large amount of images is replaced by kernel samples, have achieved less pre-training demand, they still require hours to train their model in advance and the performance is determined by the training samples. 
To reduce the kernel prior dependency, this work designs a MCMC simulation on random Gaussian distributions to provide an approximation for the blur kernel, while neither pre-training nor labeled data is necessary, and thus improve its generalization-ability towards different degradation categories.

\subsection{Meta-learning-based Non-convex Optimization}
 
Differ to the conventional deep learning methods that focus on optimizing each task exhaustedly, meta-learning methods \cite{li2020continuous, chen2021multiagent, xia2021meta, xia2022metalearning} prefer to leverage the mutual experience of solving different tasks to obtain good generalization-ability and adaptability for the trained model. For example, Li \textit{et al}. \cite{li2020boosting} propose a metric-based meta-learning that shares a hierarchical optimization structure to handle the task-specific information and task-across knowledge. 
Recently, Xia \textit{et al}. \cite{xia2022metalearning} introduce randomness perturbations with gradient descent algorithm principle, which is realized by meta-learning the optimization trajectory during iterations. 
Inspired by \cite{xia2022metalearning}, Yang \textit{et al}. \cite{Yang2022ALearning} further extend the idea of introducing random perturbations for better convergence performance in a plug-and-play fashion on solving another formous non-convex and ill-posed weighted sum rate problem in wireless communication society, and attains state-of-the-art performance with competitive computational complexity.
In a nutshell, these approaches aim to balance the to allow better convergence in the non-convex problems. Motivated by this idea, this paper employs a meta-learning scheme for solving the SISR problem, hoping to achieve better convergence in the blind SISR problem.

\section{Notation and Problem Statement}\label{sec:Preliminary}
\subsection{Notation}
We denote vectors and matrices in bold letters.
The superscript represents the index of the iteration, while the subscript indicates the practical meaning of variables. 
We denote the neural network model by $G(\cdot)$ with parameters $\bm{\phi}$. 
In the rest of this paper, we define variables with $*$ on superscript as the well-optimized results. 
For the degradation model, let $\bm{y}$ denote the down-sampled LR image, $\bm{x}$ denotes the HR image, $\otimes$ indicates the convolution operation, $\downarrow _s$ denotes the down-sampling operation with scale factor $s$, and $\bm{k}$ denotes the blur kernel. We define $\bm{n} \sim  \mathcal{N}(0,\sigma^2)$ as a zero-mean white Gaussian noise with variance $\sigma^2$. Let $\|\cdot\|_F$ denote the Frobenius norm.
\subsection{Model-based Degradation Model}
The fundamental degradation model of image super-resolution \cite{elad1997restoration, farsiu2004advances, liu2013bayesian} is typically formulated as follows
\begin{equation}
\bm{y}=(\bm{x}\otimes \bm{k})\downarrow _s+\bm{n}. \label{eq:degradation model}
\end{equation} 
Following the maximum a posteriori (MAP) framework, the blind SISR problem can be formulated as a MAP problem:
\begin{equation} \label{eq:BSR}
\;\underset{\bm{x},\bm{k}}{\max} \; p(\bm{y}|\bm{x},\bm{k}) p(\bm{x}) p(\bm{k}).\\
\end{equation}
In Eq. (\ref{eq:BSR}), $p(\bm{y}|\bm{x},\bm{k})$ denotes the likelihood of the observed LR image, $p(\bm{k})$ denotes the kernel prior, and $p(\bm{x})$ is the image prior. When optimizing the log-likelihood of \eqref{eq:BSR}, the problem can also be expressed as 
\begin{equation}
\arg\underset{\bm{x},\bm{k}}{\min}\;\|\bm{y}-(\bm{x}\otimes \bm{k})\downarrow _s\|_F^2+\lambda\Phi(\bm{x})+\beta\Omega(\bm{k}),\label{eq:BSR1}
\end{equation}
where $\Phi(\bm{x})$ and $\Omega(\bm{k})$ are the prior information functions with weights $\lambda$ and $\beta$, respectively. 
In the model-based methods, a projected alternation minimization (PAM) \cite{perrone2015clearer} based iterative optimization between $\bm{x}$ and $\bm{k}$ is typically applied to solve the problem (\ref{eq:BSR1}) in the following form
\begin{numcases}
\;\bm{x}^*= \arg \underset{\bm{x}}{\min} \;\|\bm{y} - (\bm{x}\otimes \bm{k})\downarrow _s\|_2^2+\lambda\Phi(\bm{x}),\\
\bm{k}^*= \arg \underset{\bm{k}}{\min} \;\|\bm{y} - (\bm{x}\otimes \bm{k})\downarrow _s \|_2^2+\beta\Omega(\bm{k}),\label{eq-AM-SR}
\end{numcases}

\begin{equation}
    \text{subject~to~}\|\bm{k}\|_1=1,~\bm{k}>0,~\bm{x}>0,\notag \\
\end{equation}
where the constraints are satisfied via projected gradient descent at each iteration step, more details of the PAM formulation can be found in \cite{perrone2015clearer}.

\subsection{Network-based Degradation Model}\label{sec:network-degradation}
\noindent
\textbf{Black-box based Framework}
Most of network-based degradation models, such as SRCNN \cite{dong2014learning}, RCAN \cite{zhang2018image} and  DASR \cite{wang2021unsupervised}, estimate HR image by an end-to-end deep network $\text{G}_{\bm{x}}(\cdot)$. Let $\bm{\phi}$ denote the parameters of the deep network, then the optimization problem (\ref{eq:BSR1}) is converted to
\begin{equation}
\;\bm{\phi}^{*}=\arg\underset{\bm{\phi}}{\min}\;\sum_{\bm{x}_{{gt}}^j,\bm{y}^j \in \mathcal{D}_{\bm{x}}}\|\bm{x}_{{gt}}^j-\text{G}(\bm{y}^j,\bm{\phi})\|_F^2, \label{eq:classical CNN training}  
 \end{equation}
where $\bm{x}^{{gt}}_j$ is the ground truth HR image paired with LR observation $\bm{y}_j$ in training dataset $\mathcal{D}_{\bm{x}}=\{\bm{x}_{{gt}}^j,\bm{y}^j\}^{J}_{j=1}$.

\noindent
\textbf{Double-DIP Framework}
Following the AM-based framework in (\ref{eq-AM-SR}),  Double-DIP framework \cite{gandelsman2019double, ren2020neural} establishes two networks $\text{G}_{\bm{x}}(\cdot)$ and $\text{G}_{\bm{k}}(\cdot)$ with parameters $\bm{\phi}_{\bm{x}}$ and $\bm{\phi}_{\bm{k}}$ to estimate HR image and blur kernel by taking fixed random noises $\bm{z}_{\bm{x}}$ and $\bm{z}_{\bm{k}}$ as input in the following form
\begin{numcases}
\;\bm{\phi}_{\bm{x}}^{*} = \arg\underset{\bm{\phi}_{\bm{x}}^{}}{\min}\;\|\bm{y}-(\text{G}_{\bm{x}}(\bm{z}_{\bm{x}}, \bm{\phi}_{\bm{x}})\otimes\bm{k})\downarrow _s\|_F^2,\\
\bm{\phi}_{\bm{k}}^{*} = \arg\underset{\bm{\phi}_{\bm{k}}^{}}{\min}\;\|\bm{y}-(\bm{x}\otimes\text{G}_{\bm{k}}(\bm{z}_{\bm{k}}, \bm{\phi}_{\bm{k}}))\downarrow _s\|_F^2.\label{eq:double-DIP}
\end{numcases}

\noindent
\textbf{FKP-DIP Framework}
Differ to Double-DIP framework, FKP-DIP \cite{liang2021flow} learns kernel prior via pre-training $\bm{\phi}_{\bm{k}}$ on a labelled kernel dataset $\mathcal{D}_{\bm{k}}$ as follows,
\begin{numcases}
\;\label{FKP-DIP}
\bm{\phi}_{\bm{k}}^{*}=\arg\underset{\bm{\phi}_{\bm{k}}}{\min}\;\sum_{{\bm{k}}_{gt}^j \in \mathcal{D}_{\bm{k}}}\|{\bm{k}}_{gt}^j-\text{G}(\bm{z}_{\bm{k}},\bm{\phi}_{\bm{k}})\|_F^2,\\
\bm{\phi}_{\bm{x}}^{*} = \arg\underset{\bm{\phi}_{\bm{x}}^{}}{\min}\;\|\bm{y}-(\text{G}_{\bm{x}}(\bm{z}_{\bm{x}}, \bm{\phi}_{\bm{x}})\otimes\bm{k})\downarrow _s\|_F^2,\\
\bm{z}_{\bm{k}}^{*} = \arg\underset{\bm{z}_{\bm{k}}^{}}{\min}\;\|\bm{y}-(\bm{x}\otimes\text{G}_{\bm{k}}(\bm{z}_{\bm{k}}, \bm{\phi}_{\bm{k}}^{*}))\downarrow _s\|_F^2,
\end{numcases}
where $\bm{k}_{{gt}}^j$ is the ground truth blur kernel, and $\bm{\phi}_{\bm{k}}^{*}$ denotes the pre-trained network parameters with respect to labeled kernel dataset $\mathcal{D}_{\bm{k}}=\{\bm{k}_{{gt}}^j\}^{J}_{j=1}$.

\begin{figure*}[ht!]
  \centering
  \includegraphics[width=0.9\linewidth]{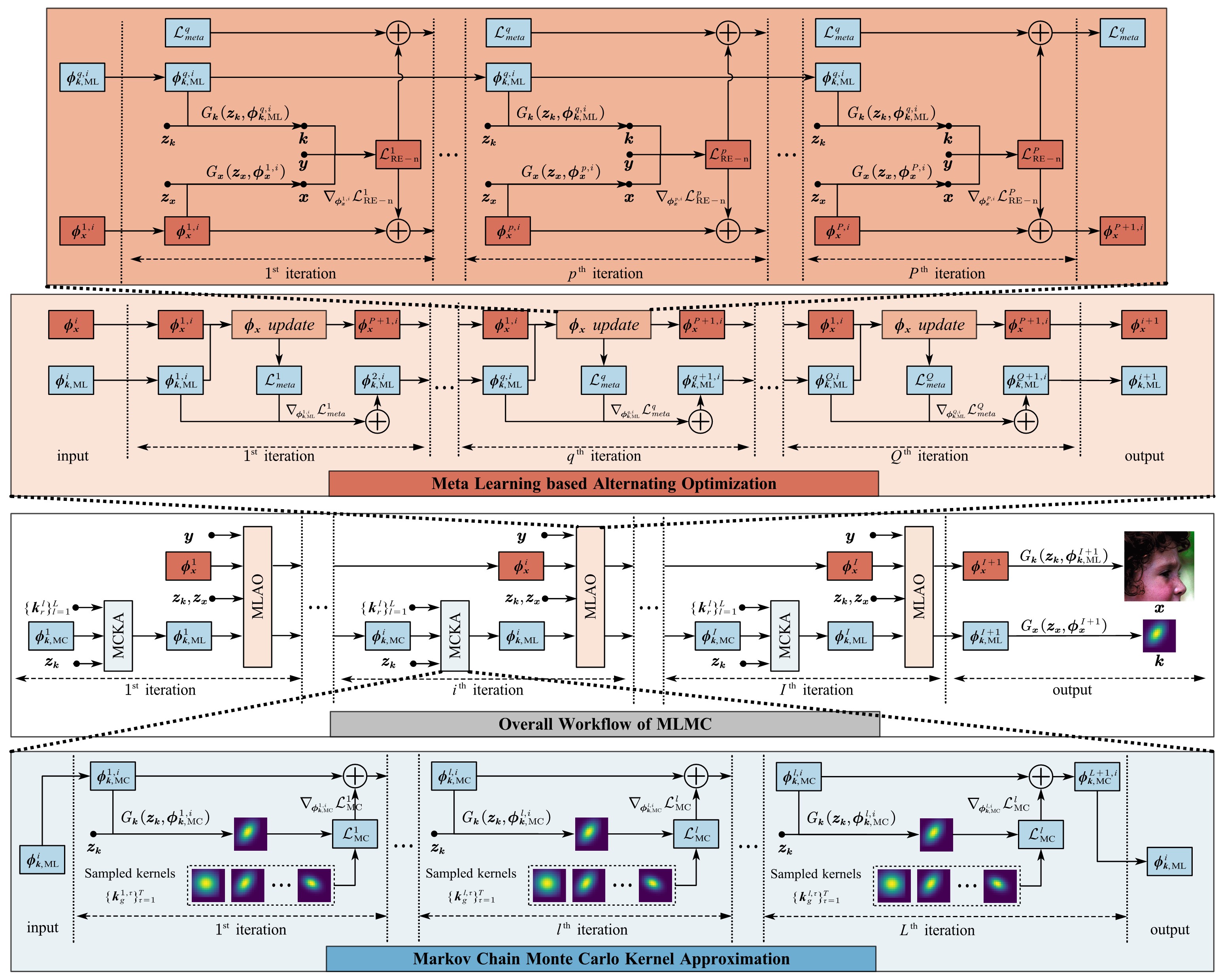}\\
  \caption{The overall framework of the proposed MLMC.}\label{fig:R1C9 overall}
\end{figure*}

\section{The Proposed MLMC Approach}\label{sec:methodology}
In this paper, a statistic formulation for a network-based degradation model is introduced, and a MLMC approach is proposed to optimize the blind SISR problem in an alternative framework, containing a MCKA stage for kernel approximation and a MLAO stage alternatives on kernel and HR image estimation.
\begin{figure*}[tpbh]
  \centering
  \includegraphics[width=0.9\linewidth]{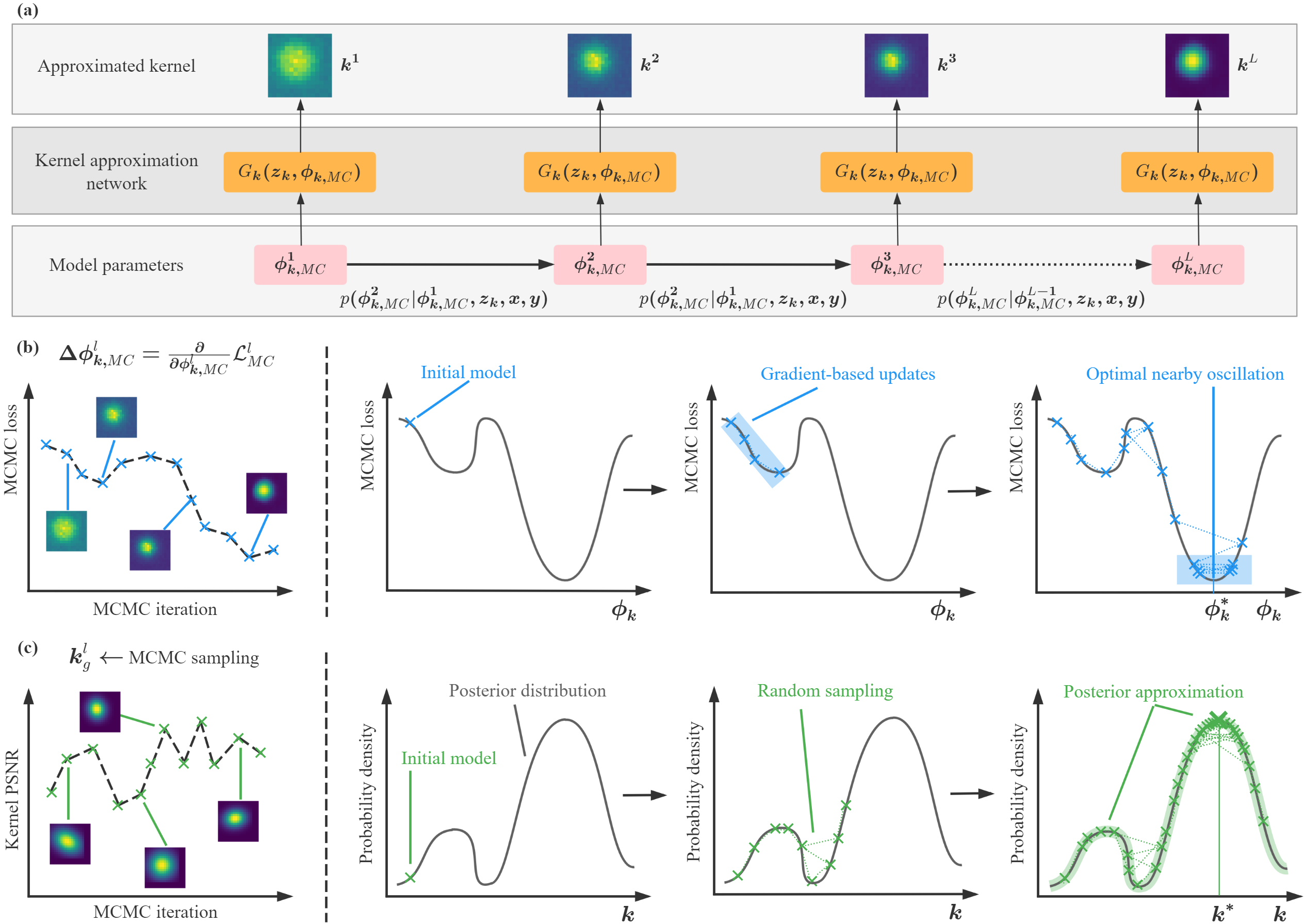}\\
  \caption{The overview of the Markov Chain Monte Carlo kernel approximation (MCKA). 
  ($\bm{a}$) The approximated blur kernel $\bm{k}^l=\text{G}_{\bm{k}}(\bm{z}_{\bm{k}},\bm{\phi}^l_{\bm{k},\textit{MC}})$ is generated from the given noise $\bm{z}_{\bm{k}}$ with a fully-connected network (FCN) $\text{G}_{\bm{k}}$, the parameters of which are iteratively updated, thereof resulting in Markov chain with possible transition$p(\bm{\phi}^{l+1}_{\bm{k},\textit{MC}}|\bm{\phi}^{l}_{\bm{k},\textit{MC}},\bm{z}_{\bm{k}},\bm{x},\bm{y})$.
    ($\bm{b}$) Left: the network parameters (blue crosses) are optimized by gradient descent-based algorithm iteratively with respect to the MCMC loss function in Eq. (\ref{eq:MCMC loss function}). The obtained approximated kernels present a trend of non-monotonically decreasing MCMC loss.
Right: the three panels show the trajectory of optimization on $\bm{\phi}^{l}_{\bm{k},\textit{MC}}$ over the geometry of MCMC loss (blue crosses). From an initial model (obtained from the last MLAO phase), the parameters $\bm{\phi}^{l}_{\bm{k},\textit{MC}}$ are updated via gradient descent-based algorithm. 
 ($\bm{c}$) Left: the MCMC simulations sample kernels (green crosses) from random Gaussian distributions with respect to a posterior that the sampled kernel should minimize the MCMC loss as well as close to the last generated sample. What stands out in this chart is the fluctuated convergence of the sampled kernel PSNR. This is caused by the posterior distribution formed by the Markov chain possible transition $p(\bm{\phi}^{l+1}_{\bm{k},\textit{MC}}|\bm{\phi}^{l}_{\bm{k},\textit{MC}},\bm{z}_{\bm{k}},\bm{x},\bm{y})$.
 Right: the three panels illustrate that sufficient MCMC sampling will lead to an approximation of posterior distribution underlying the minimized MCMC loss with respect to $\bm{k}^*$. The random samples (green crosses) are distributed on posterior distribution with the guidance of probability density from LR reconstruction loss. Those kernels with less MCMC loss will be sampled with higher probability along with the MCMC process. We note that $\bm{k}^*$ is the generated blur kernel with respect to $\bm{\phi}_{\bm{k}}^*$. 
  }\label{fig:MCKA}
\end{figure*}

\subsection{Problem Formulation for Network-based Degradation Model}\label{sec:MLMC overview} 
Following the concept of the network-based degradation model that learns to optimize HR image and blur kernel via network-behavior, 
the optimization of $\bm{x}$ and $\bm{k}$ are converted into the optimization of parameters $\bm{\phi}_{\bm{x}}$ and $\bm{\phi}_{\bm{k}}$. 

Mathematically, according to the Bayes theorem, we have 
\begin{equation}
p(\bm{x})=p(\bm{x}|\bm{\phi}_{\bm{x}})p(\bm{\phi}_{\bm{x}})/p(\bm{\phi}_{\bm{x}}|\bm{x}),\label{eq:Bayes1}
\end{equation}
\begin{equation}
p(\bm{k})=p(\bm{k}|\bm{\phi}_{\bm{k}})p(\bm{\phi}_{\bm{k}})/p(\bm{\phi}_{\bm{k}}|\bm{k}).\label{eq:Bayes2}
\end{equation}
Then, the primary MAP problem (\ref{eq:BSR}) can be reformulated as 
\begin{align}
\;\max_{\bm{\phi}_{\bm{x}},\bm{\phi}_{\bm{k}}}\;\frac{p(\bm{y}|\bm{x},\bm{k})p(\bm{x}|\bm{\phi}_{\bm{x}})p(\bm{\phi}_{\bm{x}})p(\bm{k}|\bm{\phi}_{\bm{k}})p(\bm{\phi}_{\bm{k}})}{p(\bm{\phi}_{\bm{x}}|\bm{x})p(\bm{\phi}_{\bm{k}}|\bm{k})}. \label{eq:MLMC Statistic1}
\end{align}
Given the fact that the obtained HR image $\bm{x}$ and $\bm{k}$ are determined by the parameters $\bm{\phi}_{\bm{x}}$ and $\bm{\phi}_{\bm{k}}$, respectively, thus $p(\bm{\phi}_{\bm{x}}|\bm{x})=1$ and $p(\bm{\phi}_{\bm{k}}|\bm{k})=1$.  Then problem \eqref{eq:MLMC Statistic1} can be further reformulated as 
\begin{equation}
\;\max_{\bm{\phi}_{\bm{x}},\bm{\phi}_{\bm{k}}}\;p(\bm{y}|\bm{x},\bm{k})p(\bm{x}|\bm{\phi}_{\bm{x}})p(\bm{k}|\bm{\phi}_{\bm{k}})p(\bm{\phi}_{\bm{x}})p(\bm{\phi}_{\bm{k}}).
\end{equation}
Given that $p(\bm{y}|\bm{x},\bm{k})p(\bm{x}|\bm{\phi}_{\bm{x}})p(\bm{k}|\bm{\phi}_{\bm{k}}) = p(\bm{y},\bm{x},\bm{k}|\bm{\phi}_{\bm{x}},\bm{\phi}_{\bm{k}})$, then we have 
\begin{equation}\label{statistic multi}
\max_{\bm{\phi}_{\bm{x}},\bm{\phi}_{\bm{k}}}\;p(\bm{y},\bm{x},\bm{k}|\bm{\phi}_{\bm{x}},\bm{\phi}_{\bm{k}})p(\bm{\phi}_{\bm{x}})p(\bm{\phi}_{\bm{k}}).
\end{equation}
When rewriting problem \eqref{statistic multi} in logarithm form, a universal statistical framework for network-based degradation model is given by 
\begin{align}
\;\max_{\bm{\phi}_{\bm{x}},\bm{\phi}_{\bm{k}}} \;\log p(\bm{y},\bm{x},\bm{k}|\bm{\phi}_{\bm{x}},\bm{\phi}_{\bm{k}}) + \log p(\bm{\phi}_{\bm{x}}) + \log p(\bm{\phi}_{\bm{k}}),\label{eq:MLMC Statistic2}
\end{align}
where the first term denotes a maximum log-likelihood estimation for parameters $\bm{\phi}_{\bm{x}}$ and $\bm{\phi}_{\bm{k}}$ with respect to the joint probability of LR observation $\bm{y}$, HR image $\bm{x}$ and blur kernel $\bm{k}$. The second and third terms in \eqref{eq:MLMC Statistic2} denote the image prior and kernel prior, respectively. 

Specifically, the first term is typically demonstrated by a minimization problem of LR image reconstruction error with respect to alternatively optimizing parameters $\bm{\phi}_{\bm{x}}$ and $\bm{\phi}_{\bm{k}}$, as what Double-DIP does in \eqref{eq:double-DIP}. The black-box-based end-to-end deep learning solutions, such as SRCNN \cite{dong2014learning}, RCAN \cite{zhang2018image} and DASR \cite{wang2021unsupervised}, are typically pre-trained on labeled HR image dataset $\mathcal{D}_{\bm{x}}$, consequently resulting in a strong image prior. These methods tend to learn an effective image prior from labeled data, referring to $p(\bm{\phi}_{\bm{x}}|\mathcal{D}_{\bm{x}})$. Meanwhile, classical image priors are easy to apply to $p(\bm{\phi}_{\bm{x}})$ in the form of regularization or constraint aligned with LR reconstruction error. In contrast, the latest blind SISR approach, FKP-DIP, tends to learn kernel prior instead of image prior in the form of trainable parameters by pre-trained on kernel dataset $\mathcal{D}_{\bm{k}}$, referring to $p(\bm{\phi}_{\bm{k}}|\mathcal{D}_{\bm{k}})$. 

On the basis of the problem \eqref{eq:MLMC Statistic2}, three developed modules for kernel prior, image prior and maximum likelihood estimation, respectively, are proposed in this paper as follows: i) In Section \ref{sec:kernel two stage}, a Markov Chain Monte Carlo simulation on random Gaussian distributions $\bm{k}_g$, referring to kernel approximation $p(\bm{\phi}_{\bm{k}}|\bm{k}_g)$, is proposed to substitute pre-training based kernel prior $p(\bm{\phi}_{\bm{k}}|\mathcal{D}_{\bm{k}})$; ii) In Section \ref{sec:image noise}, a hyper-Laplacian prior is adopted to improve the denoising performance, referring to $p(\bm{\phi}_{\bm{x}}|\bm{n})$;
iii) In Section \ref{sec:meta-learning}, a meta-learning based alternating optimization strategy is incorporated to optimize parameters $\bm{\phi}_{\bm{x}}$ and $\bm{\phi}_{\bm{k}}$ to improve the convergence performance of solving the maximum likelihood estimation for $p(\bm{y},\bm{x},\bm{k}|\bm{\phi}_{\bm{x}},\bm{\phi}_{\bm{k}})$. 
Detailed demonstrations of each part are given in the following subsections. 

\begin{algorithm}[t!]
\SetAlgoLined
\textbf{Input}: $\bm{y}$.\\
\textbf{Initialized}: 
$\bm{\phi}_{\bm{x}}^{1}$, $\bm{\phi}_{\bm{k},\textit{ML}}^{1}$, $\bm{\phi}_{\bm{k},\textit{MC}}^{1}$, $\bm{z}_{\bm{x}}$, $\bm{z}_{\bm{k}}$.
\\

\For{$i \gets 1, 2, \ldots, I$}{

            \% Markov Chain Monte Carlo Kernel Approximation\\
            $\bm{\phi}^{i}_{\bm{k},\textit{ML}}=f_{\textit{MCKA}}(\{\bm{k}_{g}^{l,\tau}\}_{\tau=1}^{T},\bm{z}_{\bm{k}},\bm{\phi}^{i}_{\bm{k},\textit{MC}})$\\


            \% Meta-learning based Alternating Optimization\\
            $\bm{\phi}^{i+1}_{\bm{x}}, \bm{\phi}^{i+1}_{\bm{k},\textit{MC}}=f_{\textit{MLAO}}(\bm{\phi}^{i}_{\bm{k},\textit{ML}},\bm{\phi}^{i}_{\bm{x}},\bm{y},\bm{z}_{\bm{x}},\bm{z}_{\bm{k}})$\\


}
\textbf{Output}: $\bm{k}=\text{G}_{\bm{k}}(\bm{z}_{\bm{k}}, \bm{\phi}_{\bm{k}, \textit{ML}}^{I})$, $\bm{x}=\text{G}_{\bm{x}}(\bm{z}_{\bm{x}}, \bm{\phi}_{\bm{x}}^{I})$ \\
\caption{\textcolor{black}{The overall framework of the proposed MLMC algorithm.}}
\label{alg1:MLMC overall}
\end{algorithm}

\textcolor{black}{
\subsection{Overall Framework}
The overall framework of the proposed MLMC is depicted in Fig. \ref{fig:R1C9 overall}. The fundamental paradigm is established on an alternating optimization between MCKA (blue) and MLAO (yellow) phases over the parameters of kernel estimator network $\bm{\phi}_{\bm{k}}$ and image restorer network $\bm{\phi}_{\bm{x}}$. Specifically, $\bm{\phi}_{\bm{k}}$ are optimized across these two phases referring to $\bm{\phi}_{\bm{k},\textit{MC}}$ and $\bm{\phi}_{\bm{k},\textit{ML}}$, while $\bm{\phi}_{\bm{x}}$ are only optimized within MLAO. The alternative optimization process between MCKA and MLAO is referred as the outer loop with index $i=1,2,\cdots,I$. At each outer loop step, there are MCKA inner loop with the index $l=1,2,\ldots,L$ and MLAO inner loop with the index $q=1,2,\ldots, Q$, respectively. In the MCKA inner loop, the parameters of kernel generator $\bm{\phi}_{\bm{k},\textit{MC}}$ are iteratively updated with respect to $L$ number of MCMC simulations on random Gaussian distributions to endow random kernel priors. In the MLAO inner loop, the alternates between kernel estimation and HR image restoration are processed for $Q$ iterations to ensure the estimation accuracy on the basis of the given LR observation. 
Mathematically, MCKA phase generates random kernel priors $\{\bm{k}_{g}^{l,\tau}\}_{\tau=1}^{T}$ to optimize parameters $\bm{\phi}_{\bm{k},\textit{MC}}$ as follows
\begin{equation}
    \bm{\phi}^{i}_{\bm{k},\textit{ML}}=f_{\textit{MCKA}}(\{\bm{k}_{g}^{l,\tau}\}_{\tau=1}^{T},\bm{z}_{\bm{k}},\bm{\phi}^{i}_{\bm{k},\textit{MC}}),
\end{equation}where $f_{\textit{MCKA}}$ denotes the MCKA optimization process, and $\bm{z}_{\bm{k}}$ is a random noise taken as input of the kernel estimator. The obtained $\bm{\phi}^{i}_{\bm{k},\textit{ML}}$ will be delivered to MLAO to provide a rational initialization of blur kernel in HR image restoration, and the parameters are optimized with respect to the LR observation:
\begin{equation}
   \bm{\phi}^{i+1}_{\bm{x}}, \bm{\phi}^{i+1}_{\bm{k},\textit{MC}}=f_{\textit{MLAO}}(\bm{\phi}^{i}_{\bm{k},\textit{ML}},\bm{\phi}^{i}_{\bm{x}},\bm{y},\bm{z}_{\bm{x}},\bm{z}_{\bm{k}}),
\end{equation} where $f_{\textit{MLAO}}$ denotes the meta-learning-based optimization for HR image restoration, and $\bm{z}_{\bm{x}}$ is a random noise input to the image restorer. 
The overall framework of the proposed MLMC is given in Algorithm \ref{alg1:MLMC overall}.
We will delineate these two phases as follows.}

\subsection{Markov Chain Monte Carlo Kernel Approximation}\label{sec:kernel two stage}
\textcolor{black}{
In this subsection, a kernel approximation method via MCMC simulations on random Gaussian distributions is proposed to provide rational kernel priors for $p(\bm{\phi}_{\bm{k}})$. This process is named MCKA and we elaborate it in terms of the MCMC formulation, the random kernel sampling process, the optimization of kernel estimator, and the comparison towards the existing kernel estimators as follows}. 

\subsubsection{The MCMC Formulation for Kernel Estimator}
The MCMC method is widely used to systematically generate random samples from distributions, e.g. Gaussian, which allows the algorithms to obtain a sample of the desired distribution by establishing a Markov chain that achieves equilibrium at the desired distribution. In light of this, we proposed CMC simulation on random Gaussian distributions to provide kernel priors, and construct a Markov chain on the parameters of the kernel generator, which are updated iteratively during the MCMC process to approximate blur kernels for blind SISR task. The proposed MCKA contains two major parts, the MCMC simulation on random Gaussian distributions and the optimization on kernel approximation network.  

Mathematically, let $l=1,2,\ldots,L$ denotes the index of MCMC step. At the $l^{th}$ MCMC step, a shallow FCN $\text{G}_{\bm{k}}(\cdot)$ with parameters $\bm{\phi}^{l}_{\bm{k},\textit{MC}}$ is applied to generate blur kernel $\bm{k}^l$ via taking a fixed random noise $\bm{z}_{\bm{k}}$ as input in the following form:
\begin{equation}\label{eq:generate k}
\bm{k}^l= \text{G}_{\bm{k}}(\bm{z}_{\bm{k}},\bm{\phi}^l_{\bm{k},\textit{MC}}).
\end{equation}
Meanwhile, let $\bm{k}_g$ denotes a blur kernel with kernel size $(2d+1)\times(2d+1)$ from random Gaussian distributions, 
\begin{equation}\label{eq:randomly generated kernel}
\bm{k}_g = p(\bm{h}|\bm{\Sigma}) = \frac{1}{2\pi|\bm{\Sigma}|^{-1/2}}\text{exp}\{(\bm{h}-\bm{h}_0)^T\bm{\Sigma}(\bm{h}-\bm{h_0})\},
\end{equation}
where $\bm{h} = $
\begin{math}[\begin{smallmatrix}
m\\n\end{smallmatrix}]
\end{math}
denotes the coordinates, $m,n \in [-d,d]$, $\bm{\Sigma}=$
\begin{math}[\begin{smallmatrix}
\sigma_1 \; \rho \\ \;\rho \;\; \sigma_2\end{smallmatrix}]
\end{math} denotes the covariance matrix, ${\sigma_1}$ and ${\sigma_2}$ are the horizontal and vertical variances, $\rho$ is the kernel additional random rotation angle, and $\bm{h}_0$ denotes the kernel centre coordinate. This indicates that the randomly generated distributions cover different kernel sizes $d$, center $\bm{h}_0$, and categories $\bm{\Sigma}$.
At the $l^{th}$ MCMC step, according to Eq. (\ref{eq:randomly generated kernel}), the MCMC sampled random kernel $\bm{k}^l_g$ is given by
\begin{equation}\label{eq:kernel integral}
\bm{k}^l_g = \int_{\bm{\Sigma}} p(\bm{h}|\bm{\Sigma}) p(\bm{\Sigma}|\bm{\phi}^{l}_{\bm{k},\textit{MC}}, \bm{z}_{\bm{k}},\bm{x},\bm{y}) d\bm{\Sigma},
\end{equation}
where the $p(\bm{\Sigma}|\bm{\phi}^{l}_{\bm{k},\textit{MC}}, \bm{z}_{\bm{k}},\bm{x},\bm{y})$ denotes the the posterior over the current parameters state $\bm{\phi}^{l}_{\bm{k},\textit{MC}}$, the given HR image $\bm{x}$ and LR image $\bm{y}$.

Then, the Markov chain is established on a sequence of generated blur kernels that are determined by the corresponding network parameters $\bm{\phi}^{1}_{\bm{k},\textit{MC}}, \bm{\phi}^{2}_{\bm{k},\textit{MC}}, \dots, \bm{\phi}^{L}_{\bm{k},\textit{MC}}$. 
The network parameters $\bm{\phi}^{l}_{\bm{k},\textit{MC}}$ are iteratively updated with respect to the Markov chain transition possibility in the following form
\begin{align}\label{eq:MCKA}
&\log p(\bm{\phi}^{l+1}_{\bm{k},\textit{MC}}, \bm{k}^l_g|\bm{\phi}^{l}_{\bm{k},\textit{MC}}, \bm{z}_{\bm{k}}, \bm{x},\bm{y}) = \notag \\  
&\log p(\bm{\phi}^{l+1}_{\bm{k},\textit{MC}}|\bm{k}^l_g, \bm{\phi}^{l}_{\bm{k},\textit{MC}}, \bm{z}_{\bm{k}}) + \log p(\bm{k}^l_g|\bm{\phi}^{l}_{\bm{k},\textit{MC}},\bm{z}_{\bm{k}},\bm{x},\bm{y}),
\end{align}
where $\log p(\bm{\phi}^{l+1}_{\bm{k},\textit{MC}}|\bm{k}^l_g, \bm{\phi}^{l}_{\bm{k},\textit{MC}}, 
\bm{z}_{\bm{k}})$ denotes that the state transition probability is determined by the sampled random kernel $\bm{k}^l_g$, $\bm{\phi}^l_{\bm{k},\textit{MC}}$ and $\bm{z}_{\bm{k}}$. $\log p(\bm{k}^l_g|\bm{\phi}^{l}_{\bm{k},\textit{MC}},\bm{z}_{\bm{k}},\bm{x},\bm{y})$ denotes the MCMC sampling posterior based on the LR image reconstruction error and the approximated blur kernel. Note that we have omitted the prior terms $p(\bm{x})$, $p(\bm{y})$ and $p(\bm{z}_{\bm{k}})$, which are fixed during the MCKA stage. Eq. (\ref{eq:MCKA}) illustrates that the proposed MCMC process is composed of two major modules: the update on $\bm{\phi}^l_{\bm{k},\textit{MC}}$, referring to the optimization on the kernel approximation network, and the random sampling for $\bm{k}^l_g$, referring to the MCMC simulation on random Gaussian distributions, respectively. 

Fig. \ref{fig:MCKA} illustrates the algorithmic principle and objective of the MCKA phase. Specifically, Fig. \ref{fig:MCKA} (a) shows that the approximated kernels are iteratively optimized based on the Markov chain update on $\bm{\phi}^l_{\bm{k},\textit{MC}}$. It can be seen that the approximated blur kernels are iteratively optimized to be a rational blur kernel with respect to the updates on $\bm{\phi}^l_{\bm{k},\textit{MC}}$. In the first column of Fig. \ref{fig:MCKA} (b), it explicitly shows that with the update number, the optimization loss of $\bm{\phi}^l_{\bm{k},\textit{MC}}$ non-monotonically decreases. This is because that the update is implemented in the way of gradient-based strategy as shown in the third column of Fig. \ref{fig:MCKA} (b). Moreover, thanks to the randomly sampled kernels in MCMC simulations, the optimization on $\bm{\phi}^l_{\bm{k},\textit{MC}}$ is able to escape from bad local optimum and converges to an equilibrium stationary mode as the fourth column of Fig. \ref{fig:MCKA} (b) presented. This is achieved by the parallel MCMC simulations that provide organized randomness to guide the optimization process in Fig. \ref{fig:MCKA} (b). 

Specifically, the randomly sampled kernels are shown in Fig. \ref{fig:MCKA} (c). The first column shows the general trend of the sampled kernels with update number. Though the MCMC simulations retain significant fluctuation on kernel PSNR, a converging trend is presented. In the second to the fourth columns, the posterior distribution with respect to the LR image reconstruction error loss is presented, where the highest probability density refers to the blur kernel $\bm{k}^{*}$ that minimizes the LR image reconstruction error. As shown in the fourth column of Fig. \ref{fig:MCKA} (c), the MCMC simulations randomly sample on the posterior distribution, and thereby realize organized randomness that leads to dense distribution around the $\bm{k}^{*}$. The fourth columns in Fig. \ref{fig:MCKA} (b) and (c) illustrate that the organized random sampling on posterior distribution allows the optimization on $\bm{\phi}^l_{\bm{k},\textit{MC}}$ escaping from bad local optimum and achieving equilibrium oscillation around the optima. We conclude that the updates on $\bm{\phi}^l_{\bm{k},\textit{MC}}$ in Fig. \ref{fig:MCKA} (b) and the random sampling from MCMC simulations in Fig. \ref{fig:MCKA} (c) are iteratively processed in MCKA iterations underlying learning from random Gaussian distributions that are organized by the MCMC based loss, therefore, attaining rational blur kernel as priors and providing better convergence performance.

\subsubsection{Random Kernel Sampling Process}
A Monte Carlo simulation on random Gaussian distributions is proposed with respect to the Markov chain on the state of kernel generator parameters $\bm{\phi}^l_{\bm{k},\textit{MC}}$. These parameters determine the approximated blur kernels, which are iteratively updated during the MCKA steps on the basis of the randomly sampled kernels and LR reconstruction error.

In practice, the $l^{th}$ sampled kernel $\bm{k}^l_g$ is nontrivial, thereof, the integral operation is replaced by the summation of sufficient sampling on the whole distribution, as the classic Monte Carlo simulation does as follows
\begin{align}\label{eq:kernel sum}
\bm{k}^l_g &\approx \sum^T_{\tau=1} p(\bm{h}|\bm{\Sigma}^{l,\tau})p(\bm{\Sigma}^{l,\tau}| \bm{\phi}^{l}_{\bm{k},\textit{MC}}, \bm{z}_{\bm{k}},\bm{x},\bm{y}) 
= \sum^T_{\tau=1} { \omega}^{l,\tau}_{} \bm{k}^{l,\tau}_g,
\end{align}
where $p(\bm{h}|\bm{\Sigma}^{l,\tau})$ can be referred to the $\tau^{th}$ Gaussian kernel $\bm{k}^{l,\tau}_g$ dominated by the covariance matrix $\bm{\Sigma}^{l,\tau}$ with $\sigma_1, \sigma_2 \in (0, \sigma_{max}]$ and $\rho \in [-\pi,\pi]$. $\sigma_{max}$ denotes the maximum variance of kernel boundary,
and $T$ is the total number of random samples in one Monte Carlo simulation.
$p(\bm{h}|\bm{\Sigma}^{l,\tau})p(\bm{\Sigma}^{l,\tau}| \bm{\phi}^{l}_{\bm{k},\textit{MC}}, \bm{z}_{\bm{k}},\bm{x},\bm{y})$ can be referred to the weight $\omega^{l,\tau}$ of the $\tau^{th}$ sampled kernel.
Different to the conventional Monte Carlo simulation that uniformly samples blur kernels from random Gaussian distributions, in this paper, the Markov chain on $\bm{\phi}^{l}_{\bm{k},\textit{MC}}$ re-weights all the sampled kernels by kernel weight ${\omega^{l,\tau}}$  in one Monte Carlo simulation as follows
\begin{align}\label{eq:MCMC weight}
{\omega^{l,\tau}} = p(\bm{k}_{g}^{l,\tau}|\bm{\phi}^{l}_{\bm{k},\textit{MC}}, \bm{z}_{\bm{k}},\bm{x},\bm{y}) \propto \frac{1}{{\nu}^{l,\tau}_{}} 
\end{align}
\begin{align}\label{eq:MCMC weight loss}
\;\nu^{l,\tau}_{} = \|\bm{y}-(&\bm{x}\otimes\bm{k}_g^{l,\tau})\downarrow _s\|_F^2, \notag\\
& + \|\text{G}_{\bm{k}}(\bm{z}_{\bm{k}},\bm{\phi}^l_{\bm{k},\textit{MC}}) - \bm{k}_g^{l,\tau})\|_F^2 + \epsilon,&
\end{align}
where $\epsilon$ is a small hyper-parameter to prevent ${\nu}^{l,\tau}_{}=0$. Eq. (\ref{eq:MCMC weight}) and Eq. (\ref{eq:MCMC weight loss}) illustrate that the randomly sampled kernels are weighted by the $\nu^{l,\tau}$ that is composed of the LR image reconstruction term and the MSE of the sampled kernel and approximated kernel at the $l^{th}$ iteration. Therefore, we claim that the randomly sampled kernels are organized to be distributed closely to the kernels that minimize the LR image reconstruction error, and the new sampling will prefer to occur in the neighbor of the last sampled kernel. The LR image reconstruction error urges the randomly sampled kernels to realize the posterior distribution as shown in Fig. \ref{fig:MCKA} (c) right, while the MSE brings a balance between steep variation on kernel PSNR and stable convergence on the obtained rational blur kernel as shown in Fig. \ref{fig:MCKA} (c) left.

\begin{figure}[t]
  \centering
  \includegraphics[width=1\linewidth]{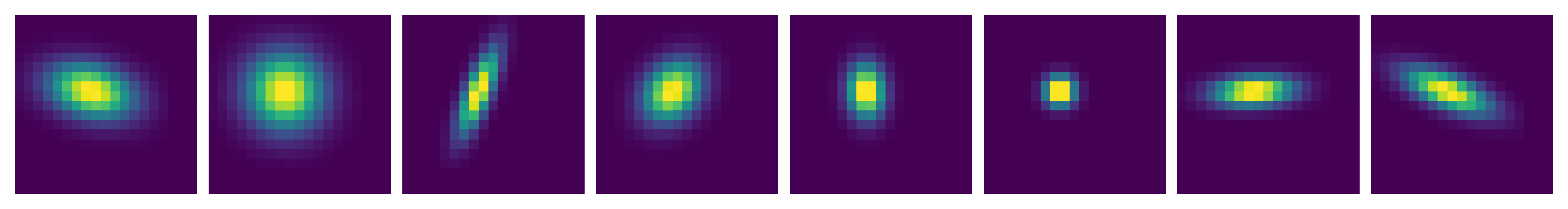}\\
  \caption{Blur kernels are randomly sampled by the Monte Carlo simulation from different Gaussian distributions, e.g., different variances ($\bm{\Sigma}$) and rotation angle ($\rho$). }\label{fig:Kernels}
  \vspace{-0.2cm}
\end{figure}

We note that, instead of exhaustively sampling on the full range of $\bm{\Sigma}$ with extremely large sampling number $T$, as the classical MCMC method does, only a few realizations are randomly sampled in one MCKA stage. In Fig. \ref{fig:Kernels}, the visualization of the randomly sampled blur kernels from different Gaussian distributions in one Monte Carlo simulation is presented. It is explicit that the variability of these sampled kernels in terms of coordinate and outline is significantly large. After the Markov chain possible transition reweights the sampled kernels, the obtained integrated kernel $\bm{k}_g^l$ is applied to optimize the parameters $\bm{\phi}^{l}_{\bm{k},\textit{MC}}$. In this way the obtained sampled kernel $\bm{k}^l_g$ can be regarded as a loose relaxation for the obtained approximated blur kernel $\bm{k}^l$, which is dominated by parameters $\bm{\phi}^{l}_{\bm{k},\textit{MC}}$, which typically suffers from the overfitting optimization during the MLAO stage even with the proposed meta-learning-based framework. The optimization on approximation network will be demonstrated next.

\subsubsection{Optimizing the Kernel.}
The parameters of kernel approximation network $\bm{\phi}_{\bm{k},\textit{MC}}$ are iteratively optimized with respect to the randomly sampled Gaussian distributions at each MCKA iteration. Specifically, at each iteration $l$, recalling the transition possible in Eq. (\ref{eq:MCKA}), the optimization on $\bm{\phi}^{l}_{\bm{k},\textit{MC}}$ can be expressed as a log-likelihood maximization problem on transition probability in the following form
\begin{align}\label{eq:KA MSE}
&{\max} \log p(\bm{\phi}^{l+1}_{\bm{k},\textit{MC}}|\bm{k}^l_g,\bm{\phi}^{l}_{\bm{k},\textit{MC}},\bm{z}_{\bm{k}},\bm{x},\bm{y}) \notag\\
=&{\max} \log p(\bm{\phi}^{l+1}_{\bm{k},\textit{MC}}|\{\bm{k}^{l,\tau}_g \}^{T}_{\tau=1},\bm{\phi}^{l}_{\bm{k},\textit{MC}},\bm{z}_{\bm{k}},\bm{x},\bm{y}),
\end{align}  
where each random sampled kernel $\bm{k}^{l,\tau}_g$ is independent and identically distributed (i.i.d.). However, each $\bm{k}^{l}_g$ entails Markov chain based re-weighting recalling to $\omega^{l,\tau}$ in Eq. (\ref{eq:MCMC weight loss}), which bridges the possible transition with the LR reconstruction in terms of $\bm{x}$ and $\bm{y}$. Therefore, the optimization problem (\ref{eq:KA MSE}) can be rewritten as  
\begin{align}
&{\max} \log \prod^T_{\tau=1} p(\bm{\phi}^{l+1}_{\bm{k},\textit{MC}}|\bm{k}^{l,\tau}_g)p(\bm{k}^{l,\tau}_g|\bm{\phi}^{l}_{\bm{k},\textit{MC}},\bm{z}_{\bm{k}},\bm{x},\bm{y}) \notag \\
=& {\min}\sum^T_{\tau=1} -\log p(\bm{\phi}^{l+1}_{\bm{k},\textit{MC}}|\bm{k}^{l,\tau}_g)p(\bm{k}^{l,\tau}_g|\bm{\phi}^{l}_{\bm{k},\textit{MC}},\bm{z}_{\bm{k}},\bm{x},\bm{y}).\label{eq:KA MSE2}
\end{align}

As demonstrated in Eq. \eqref{eq:MCMC weight}, $\omega^{l,\tau}=p(\bm{k}^{l,\tau}_g|\bm{\phi}^{l}_{\bm{k},\textit{MC}},\bm{z}_{\bm{k}},\bm{x},\bm{y})$.
Then, the MCMC loss $\mathcal{L}_\textit{MC}^{l}$ is given by
\begin{equation}\label{eq:MCMC loss function}
\underset{\bm{\phi}^{l}_{\bm{k},\textit{MC}}}{\min}\mathcal{L}^l_\textit{MC} = \sum_{\tau=1}^{T} \omega^{l,\tau}\|\text{G}_{\bm{k}}(\bm{z}_{\bm{k}},\bm{\phi}^l_{\bm{k},\textit{MC}}) - \bm{k}^{l,\tau}_g) \|_F^2.
\end{equation}
Eq. (\ref{eq:MCMC loss function}) elaborates that the parameters $\bm{\phi}^{l}_{\bm{k},\textit{MC}}$ are optimized with respect to the MSE between the approximated blur kernels and the randomly sampled kernel, underlying the organized by the MCMC weight $\omega^{l,\tau}$. The parameters are optimized by the Adam \cite{kingma2014adam} optimizer:
\begin{equation}\label{eq:KA meta-update}
\bm{\phi}_{\bm{k},\textit{MC}}^{l+1} = \bm{\phi}_{\bm{k},\textit{MC}}^{l}-\gamma_\textit{MC}^{l}\cdot\mathrm{Adam}\left(\frac{\partial}{\partial\bm{\phi}_{\bm{k},\textit{MC}}^l} \mathcal{L}_\textit{MC}^l\right),
\end{equation} where $\gamma_\textit{MC}^{l}$ denotes the learning rate.

\begin{algorithm}[t!]
\SetAlgoLined
\textbf{Input}: $\bm{y}$, $\bm{z}_{\bm{k}}$, $\bm{\phi}^{i}_{\bm{k},\textit{MC}}$.\\


            \% Markov Chain Monte Carlo kernel approximation
            
            $\bm{\phi}^{1, i}_{\bm{k},\textit{MC}} \gets \bm{\phi}^{i}_{\bm{k},\textit{MC}}$ 
            
            Sample random kernels $\{\bm{k}_g^{l,\tau}\}^T_{\tau=1}$ via MC sampling
            
            \For{$l \gets 1, 2, \ldots, L$}{
                \For{$\tau \gets 1, 2, \ldots, T$}{
                 $\nu^{l,\tau}_{} = \|\bm{y}-(\bm{x}\otimes\bm{k}_g^{l,\tau})\downarrow _s\|_F^2 + $\\ $\;\;\;\;\;\;\;\;\;\;\;\;\;\;\;\;\;\;\;\;\;\;\;\;\;\;\|\text{G}_{\bm{k}}(\bm{z}_{\bm{k}},\bm{\phi}^l_{\bm{k},\textit{MC}}) - \bm{k}_g^{l,\tau})\|_F^2 + \epsilon$\\
                 $\omega^{l,\tau} = \frac{1}{{\nu}^{l,\tau}_{}}$
                }
                $\mathcal{L}_\textit{MC}^l = \sum_{\tau=1}^{T} {\omega^{l,\tau}} \|\text{G}_{\bm{k}}(\bm{z}_{\bm{k}}, \bm{\phi}_{\bm{k},\textit{MC}}^{l,i})-{\bm{k}_g(\bm{\Sigma}^\tau)}) \|_F^2$\\
            $\bm{\phi}_{\bm{k}, \textit{MC}}^{l+1,i} = \bm{\phi}_{\bm{k}, \textit{MC}}^{l,i}-\gamma_\textit{MC}^{l,i} \mathrm{Adam}(\frac{\partial}{\partial\bm{\phi}_{\bm{k}, \textit{MC}}^{l,i}} \mathcal{L}_\textit{MC}^l)$
            } 
            $\bm{\phi}_{\bm{k}, \textit{ML}}^{i} \gets \bm{\phi}_{\bm{k}, \textit{MC}}^{L,i}$


\textbf{Output}: $\bm{\phi}_{\bm{k}, \textit{ML}}^{1,i}$ 
\caption{\textcolor{black}{The work flow of the MCKA phase.}}
\label{alg2:MCKA}
\end{algorithm}

We note that the optimization on the kernel approximation network is implemented in a way of plug-and-play fashion, that is ``training while solving" the blind SISR task, instead of training in advance. The whole MCKA stage alternatively processes the MCMC simulations for rational blur kernels and optimizes the kernel approximation network via learning from organized randomness.
There are two major contributions of enrolling the proposed MCKA procedure in solving the blind SISR problem. On the one hand, the MCMC simulations provide a loose but rational kernel prior for blur kernel estimation, while no pre-training procedure and labeled data are needed. On the other hand, the MCKA phase periodically brings random disturbance to the convergence of the parameters $\bm{\phi}_{\bm{k}}$, which are alternatively optimized with respect to the LR image reconstruction error loss in the MLAO phase. Therefore, it prevents the optimization of the parameters $\bm{\phi}_{\bm{k}}$ converging to bad local modes due to the intrinsic non-convexity and ill-posedness. We note that the randomnesses learned in MCKA will approximately converge to a desired distribution via the MCMC behavior in terms of minimizing the LR image reconstruction error. Thus, the additive disturbance will not lead to a significant deviation that is detrimental to the convergence of solving the blind SISR problem.

\textcolor{black}{The overall data flow of the MCKA phase is given in Algorithm \ref{alg2:MCKA}. It should be also noted that we set $L$ and $T$ to be small for the following two reasons: i) The obtained kernel priors are flexible, and can be easily refined by the LR image reconstruction error; ii) The implementation is less time-consuming and computationally less demanding, ensuring negligible runtime cost and memory usage.
Consequently, the MCKA is implemented in a \textcolor{black}{plug-and-play manner} and is combined with a meta-learning framework to resolve the blind SISR problem in the next section. }


\subsubsection{\textcolor{black}{Kernel Estimator Comparisons}}

\textcolor{black}{
As aforementioned in Section \ref{sec:related work}, the existing kernel estimators can be roughly divided into two categories: designing kernel model in a handcrafted way \cite{perrone2015clearer, ren2020neural, Yue2022blind} or estimating kernel via networks \cite{gandelsman2019double, liang2021flow, liang2021mutual}. 
We compare the most related existing kernel estimators as follows:
\begin{itemize}
    \item \emph{Deep-learning methods}: More recent deep learning methods also investigate to design specific modules on the basis of kernel prior knowledge, such as implicit degradation modeling \cite{xia2022knowledge}, shuffled degradation processes \cite{zhang2021designing}, and varying degradations for multiple patches \cite{liang2021mutual}. In a nutshell, these methods tend to estimate kernels via learning on labeled HR images with independent modeling on kernel priors. In contrast, the MLMC strives to learn kernel priors from randomness in an unsupervised manner and follows a plug-and-play fashion. 
    \item \emph{Double-DIP}: Double-DIP adopts a DIP-like network without pre-training for kernel estimation. The adopted network is simply optimized with respect to LR image restoration error and no specific kernel priors are available, thus the performance is not satisfactory in most Gaussian and motion kernel cases. Then, the network utilized in MLMC has two differences compared to Double-DIP: i) the network architecture is much more lightweight; ii) the parameters are optimized on the basis of random kernel priors along with the LR image restoration error.
    \item \emph{FKP-DIP}: The kernel estimator FKP module is pre-trained in a supervised way with sufficient kernel data. This training process typically requests hours and is necessary when being applied to different kernel categories. In contrast, the kernel learning module in MLMC is plug-and-play, allowing stronger flexibility for varying kernel categories.
    \item \emph{BSRDM \cite{Yue2022blind}}: In BSRDM, an explicit two-dimensional Gaussian model is employed to estimate the Gaussian kernel, whose parameters (mean and variance) are optimized via a gradient-based algorithm along with the LR image reconstruction error through the entire solution iterations. In contrast, the kernel estimator in the proposed MLMC only formulates the kernel distribution, allowing a significantly larger scope for kernel prior learning, resulting in more dynamic and flexible kernel priors. Besides, the MLMC kernel estimator is trained on the basis of learned kernel priors and LR observation, ensuring better convergence performance than BSRDM. Moreover, the MLMC can be applied to non-Gaussian kernels, such as the motion kernel. This highlights its better generalization ability and flexibility in practice.
\end{itemize}
In a nutshell, comparing to the existing kernel estimators, the proposed MLMC combines the merits of model-based and learning-based methods, realizing a training while solving paradigm for kernel estimation without requests on labeled dataset and training before applying. We claim that the random kernel-prior-learning based MCKA is the main technical contribution of the proposed MLMC and the fundamental discrepancy towards the existing kernel estimators.
}

\subsection{Meta-learning based Alternating Optimization}\label{sec:meta-learning}

Despite rational kernel priors, referring to $p(\bm{\phi}_{\bm{k}})$, can be gained from the MCKA phase, the primary blind SISR problem with respect to $p(\bm{y},\bm{x},\bm{k}|\bm{\phi}_{\bm{x}},\bm{\phi}_{\bm{k}})$ retains high non-convexity and ill-posedness, therefore, being ease of converging to bad local optimums, especially for the kernel estimation. 
Inspired by \cite{xia2022metalearning, Yang2022ALearning}, a meta-learning-based alternating optimization is designed to optimize $\bm{\phi}_{\bm{x}}$ and $\bm{\phi}_{\bm{k}}$ from the observed LR image $\bm{y}$, the whole process is named MLAO phase and elaborated as follows.
\subsubsection{HR Image restoration Optimization}
As aforementioned in Section \ref{sec:network-degradation}, the network-based blind SISR problem \eqref{eq:double-DIP} is typically solved in an AM-based fashion between the kernel estimation and image restoration sub-problems, which alternatively estimate the HR image $\bm{x}$ and blur kernel $\bm{k}$ via networks $\text{G}_{\bm{x}}(\bm{z}_{\bm{x}}, \bm{\phi}_{\bm{x}})$  and $\text{G}_{\bm{k}}(\bm{z}_{\bm{k}}, \bm{\phi}_{\bm{k}})$, respectively. 
However, the intermediate solutions from each sub-problem in \eqref{eq:double-DIP} typically contain significant noise, and the minimization behavior for each sub-problem using only first-order information may not necessarily lead to benign solutions in the view of global convergence. 
Therefore, a meta-learning scheme is proposed to learn a less-greedy and adaptive updating rule for better convergence performance in terms of kernel estimation.

Mathematically, let $p=1,2,\ldots,P$ denotes the SISR iteration index, and $q=1,2,\ldots, Q$ represents the index of the meta-update on the kernel generator. Then, at the $p^{th}$ iteration, the LR image reconstruction error is given by 
\begin{equation}\label{eq:RE x loss}
\mathcal{L}_\textit{RE}^p = \|\bm{y}-(\bm{x}^p\otimes \bm{k}^q)\downarrow _s\|_F^2,
\end{equation}
where $\bm{k}^q=\text{G}_{\bm{k}}(\bm{z}_{\bm{k}}, \bm{\phi}_{\bm{k},\textit{ML}}^{q})$ and $\bm{x}^p=\text{G}_{\bm{x}}(\bm{z}_{\bm{x}}, \bm{\phi}_{\bm{x}}^{p})$. 
Instead of exhaustively optimize each sub-problem with respect to minimizing  $\mathcal{L}_\textit{RE}^p$, the kernel estimation iterations are processed in a meta-learning manner. We tend to minimize the leveraged reconstruction errors over $P$  iterations, denoted by the meta-loss $\mathcal{L}_\textit{ML}^{q}$, to update the parameters $\bm{\phi}_{\bm{k},\textit{ML}}^{q}$ for $Q$ times as follows
\begin{numcases}
 \;\mathcal{L}_\textit{ML}^q = \frac{1}{P}\sum_{p=1}^{P} \omega^p \mathcal{L}_\textit{RE}^p,\\
\bm{\phi}_{\bm{k},\textit{ML}}^{q+1} = \bm{\phi}_{\bm{k},\textit{ML}}^{q}-\gamma_\textit{ML}^{q}\cdot\mathrm{Adam}\left(\frac{\partial}{\partial\bm{\phi}_{\bm{k},\textit{ML}}^q} \mathcal{L}_\textit{ML}^q \right),
\end{numcases}
where $\gamma_\textit{ML}^{q}$ denotes the learning rate at the $q^{th}$ meta-update step and $\omega^p$ is the weight of $\mathcal{L}_\textit{RE}^p$ at iteration $p$. In this instance, the parameters $\bm{\phi}_{\bm{k},\textit{ML}}^{q}$ are no longer optimized via minimizing each individual reconstruction error but the accumulated losses $\mathcal{L}_\textit{ML}^q$, which results in a less-greedy but more adaptive optimization strategy. As illustrated in \cite{xia2022metalearning}, $\mathcal{L}_\textit{ML}^q$ essentially leads to a better optimum, as it optimizes each individual sub-problem in-exhaustively by globally learning the mutual knowledge of the optimization strategy on partial reconstruction error trajectories across iterations. In this way, the optimization strategy for the blur kernel estimation is endowed with a relaxation, which allows a non-optimal solution at the $p^{th}$ sub-problem for $\bm{\phi}_{\bm{k}}$ during the estimation iterations but is capacity of converging to better optimum in the view of $P$ iterations.

Meanwhile, the image restoration sub-problem is optimally solved at each iteration on the basis of the hyper-Laplacian image prior demonstrated in Section \ref{sec:image noise}, which updates the parameters $\bm{\phi}_{\bm{x}}$ as follows,
\begin{equation}
\bm{\phi}_{\bm{x}}^{p+1} = \bm{\phi}_{\bm{x}}^{p}-\gamma_{\bm{x}}^{p}\cdot\mathrm{Adam}\left(\frac{\partial}{\partial\bm{\phi}_{\bm{x}}^p} \mathcal{L}_\textit{RE-n}^p\right),
\end{equation}
where $\gamma_{\bm{x}}^{p}$ denotes the learning rate. We note that the adopted network $\text{G}_{\bm{x}}$ is a well-designed image restorer with substantiated performance in the literature \cite{ulyanov2018deep,ren2020neural,liang2021flow}. Hence, the vanilla iterative minimization is sufficiently satisfactory to realize good HR image reconstruction when the blur kernel is obtained through the meta-learning-based optimization strategy. 

\textcolor{black}{
In this stage, the overview of the MLAO phase is depicted in the yellow box in Fig. \ref{fig:R1C9 overall} and the work flow is given in Algorithm \ref{alg3:MLAO}. Explicitly, the whole MLAO stage tends to refine the approximated kernel based on the meta-loss $\mathcal{L}_\textit{ML}^q$, as such $\bm{\phi}_{\bm{k},\textit{ML}}^{q}$ are optimized on the basis of the observed LR image $\bm{y}$ via a less-greedy and more adaptive optimization strategy. We note that the incorporated meta-learning approach on kernel estimation regards each kernel estimation sub-problem as a training sample, and meta-learn across kernel estimation iterations to extract the mutual knowledge of solving a set of sub-problems. In this way, the learned update rule becomes more flexible and non-monotonically, therefore, being able to prevent trapping into bad local optimum.}

\begin{algorithm}[t!]
\setlength{\baselineskip}{14pt}
\SetAlgoLined
\textbf{Input}: $\bm{\phi}^{i}_{\bm{k},\textit{ML}},\bm{\phi}^{i}_{\bm{x}},\bm{y},\bm{z}_{\bm{x}},\bm{z}_{\bm{k}}$.\\

            \% Meta-learning based alternating optimization (MLAO)

            $\bm{\phi}^{1,i}_{\bm{k},\textit{ML}} \gets \bm{\phi}^{i}_{\bm{k},\textit{ML}}, \bm{\phi}^{1,i}_{\bm{x}} \gets \bm{\phi}^{i}_{\bm{x}}$
            
        \For{$q \gets 1, 2, \ldots, Q$}{
            $\bm{k}^q=\text{G}_{k}(\bm{z}_{\bm{k}}, \bm{\phi}_{\bm{k}, \textit{ML}}^{q,i})$\\
            \% Image network update \\
            \For{$p \gets 1, 2, \ldots, P$}{
                $\bm{x}^{p}=\text{G}_{\bm{x}}(\bm{z}_{\bm{x}},\bm{\phi}_{\bm{x}}^{p,i})$\\
                $\sigma^2 = \frac{1}{h\times w} \sum_{j\in N} \{{y}_j- \left[(\bm{x}^{p}\otimes \bm{k}^q)\downarrow _s\right]_j\}^2$\\   
                $\mathcal{L}_\textit{RE-n}^p = \frac{1}{\sigma^2} \|\bm{y}-(\bm{x}^{p}\otimes \bm{k}^q)\downarrow _s\|_F^2+ $\\ $\;\;\;\;\;\;\;\;\;\;\;\;\;\;\;\;\;\;\;\;\;\;\;\;\;\;\;\;\;\;\;\;\;\;\;\;\rho\sum^{2}_{c=1}\left(\|f_c\otimes\bm{x}^{p}\|_F^2\right)^\eta $\\
                $\bm{\phi}_{\bm{x}}^{p+1,i} \gets \bm{\phi}_{\bm{x}}^{p,i}-\gamma_{\bm{x}}^{p}\cdot\mathrm{Adam}(\frac{\partial}{\partial\bm{\phi}_{\bm{x}}^{p,i}}\mathcal{L}_\textit{RE-n}^p)$
            }
            $\bm{\phi}_{\bm{x}}^{1,i} \gets \bm{\phi}_{\bm{x}}^{P,i}$\\
            \% Kernel network update \\
            $\mathcal{L}_\textit{ML}^q =\frac{1}{P} \sum_{p=1}^{P} \omega_{}^p \mathcal{L}_\textit{RE-n}^p$\\
            $\bm{\phi}_{\bm{k},\textit{ML}}^{q+1,i} \gets \bm{\phi}_{\bm{k},\textit{ML}}^{q,i} -\gamma_\textit{ML}^{q}\cdot\mathrm{Adam}(\frac{\partial}{\partial\bm{\phi}_{\bm{k},\textit{ML}}^{q,i}}\mathcal{L}_\textit{ML}^q)$
        }
    $\bm{\phi}_{\bm{x}}^{i+1} \gets \bm{\phi}_{\bm{x}}^{P,i}$, $\bm{\phi}_{\bm{k}, \textit{MC}}^{i+1} \gets \bm{\phi}_{\bm{k}, \textit{ML}}^{Q,i}$

\textbf{Output}: $\bm{\phi}^{i+1}_{\bm{x}}, \bm{\phi}^{i+1}_{\bm{k},\textit{MC}}$ 
\caption{\textcolor{black}{The work flow of the MLAO phase.}}
\label{alg3:MLAO}
\end{algorithm}

\subsubsection{Image Noise Optimization}\label{sec:image noise}
In this subsection, we further propose an HR image restoration formulation for the noise scenario. A hyper-Laplacian image prior is incorporated with the HR image restoration for $p(\bm{\phi}_{\bm{x}})$.  
We formulate an independent and identically distributed (i.i.d) additive white Gaussian noise (AWGN) model for the fundamental degradation model as follows,
\begin{equation}\label{eq:image noise}
\;\bm{y} \sim \mathcal{N}(\bm{y}|(\bm{x}\otimes\bm{k})\downarrow _s,\sigma^2),
\end{equation}
where $\sigma$ denotes the covariance of the noise distribution. 
Similar to the previous works \cite{ulyanov2018deep, ren2020neural, liang2021flow, Yue2022blind}, a CNN-based network $\text{G}_{\bm{x}}(\cdot)$ is established to estimate the HR image as formulated in Eq. (\ref{eq:double-DIP}).
As demonstrated in \cite{ulyanov2018deep}, $\text{G}_{\bm{x}}(\cdot)$ is typically fragile to the image noise, thus leading to overfitting to bad local optimums.
A hyper-Laplacian prior \cite{krishnan2009fast} is implemented to constrain the AWGN of the estimated HR image following the proposed statistic framework as follows,
\begin{equation}\label{eq:AWGN}
(\bm{\phi}_{\bm{x}}, \bm{z}_{\bm{x}}) \sim p(\bm{\phi}_{\bm{x}}, \bm{z}_{\bm{x}}) = p(\bm{\phi}_{\bm{x}}|\bm{z}_{\bm{x}})p(\bm{z}_{\bm{x}}), \\
\end{equation}
\begin{equation}
p(\bm{\phi}_{\bm{x}}|\bm{z}_{\bm{x}}) \propto \text{exp}\left(-\rho\sum^{2}_{c=1} 
\left( \|f_c\otimes\text{G}_{\bm{x}}(\bm{z}_{\bm{x}}, \bm{\phi}_{\bm{x}})\|_F^2\right)^\eta \right),
\end{equation}
\begin{equation}
p(\bm{z}_{\bm{x}}) = \mathcal{N}(0,\sigma^2),
\end{equation}
where $\{f_c\}^{2}_{c=1}$ represent the gradient filters along the horizontal and vertical directions, $\rho$ and $\eta$ denote the hyper-parameters.
Equivalently, the parameters of the image estimator $\bm{\phi}_{\bm{x}}$ are optimized by the loss function as follow,
\begin{align}
\underset{\bm{\phi}_{\bm{x}}}{\min}\;\mathcal{L}_\textit{RE-n} = \frac{1}{\sigma^2} \|\bm{y}-(\text{G}_{\bm{x}}(\bm{z}_{\bm{x}}, \bm{\phi}_{\bm{x}})\otimes \bm{k})\downarrow _s \|_F^2 \notag \\
+ \rho\sum^{2}_{c=1}\left(\|f_c\otimes\text{G}_{\bm{x}}(\bm{z}_{\bm{x}}, \bm{\phi}_{\bm{x}})\|_F^2\right)^\eta. \label{eq:image noise loss}
\end{align}
The noise variance $\sigma^2$ is typically given by
\begin{equation}\label{eq:noise update}
\sigma^2 = \frac{1}{h\times w} \sum_{j\in N} \{{y}_j-\left[(\text{G}_{\bm{x}}(\bm{z}_{\bm{x}}, \bm{\phi}_{\bm{x}})\otimes \bm{k})\downarrow _s\right]_j\}^2,
\end{equation}
where $h$ and $w$ denote the image height and width, respectively, and the number of image pixels $N=h\times w$, $j$ denotes the $j^{th}$ pixel of the image.

\textcolor{black}{
\subsection{Network-level Langevin Dynamics}}
\textcolor{black}{
The vanilla Langevin dynamics \cite{welling2011bayesian} is proposed to improve the performance of gradient descent-based optimization algorithms for variable update, which can be typically formulated as follows,
\begin{equation}\label{eq:langevin}
\bm{z}^{t+1} = \bm{z}^{t} + \frac{\delta^2}{2} \cdot \frac{\partial \log p( \bm{z}^{t}|\bm{y})}{\partial \bm{z}^t} + \delta \cdot \zeta_{\bm{n}},
\end{equation}
where $t$ denotes the index of update step, $\bm{z}$ denotes the variable to be optimized, $\bm{y}$ denotes the observation data, $\delta$ denotes the optimization step size. $\log p_{}( \bm{z}^{t}|\bm{y})$ demonstrates the standard optimization on the basis of the task-specific data, and $\zeta_{\bm{n}}$ is a random noise (e.g., zero-mean Gaussian noise). 
When $t \rightarrow \infty$ and $\delta^2 \rightarrow 0$, Langevin dynamics realizes a sampling from the task-specific data posterior $p(\bm{z}^{t}|\bm{y})$ to optimize $\bm{z}^{t+1}$.
Theoretically, this is achieved by the additive noise $\zeta_{\bm{n}}$ that correlated with the posterior $p(\bm{z}^{t}|\bm{y})$.}

\textcolor{black}{Different to the Langevin dynamics in BSRDM, the proposed MLMC tends to resolve an optimization problem of network parameters update. Following the concept of improving the optimization convergence performance via incorporating random fluctuation, the white Gaussian noise should be re-designed to fit the network-based optimization process. A natural idea is to incorporate random samples into the network training process to achieve a similar effect of Langevin dynamics. 
In light of this concept, in the proposed MLMC, we proposed a novel network-level Langevin dynamics optimization strategy that is designed for network parameters optimization as follows
\begin{align}\label{eq:network-langevin}
\bm{\phi}_{\bm{k}}^{\textit{new}} = \bm{\phi}_{\bm{k}}^{\textit{old}} + \gamma_{\textit{ML}} \cdot \frac{\partial \log p_{\textit{MLAO}}(\bm{\phi}_{\bm{k}}^{\textit{old}}|\bm{y})}{\partial \bm{\phi}_{\bm{k}}^{\textit{old}}} + \gamma_\textit{MC}^{} \cdot \zeta_{\bm{k}},
\end{align}
where $\bm{\zeta}_{\bm{k}}=\frac{\partial \log p_{\textit{MLAO}}(\bm{\phi}_{\bm{k}}^{\textit{old}}|\bm{y})}{\partial \bm{\phi}_{\bm{k}}^{\textit{old}}}$ denotes the gradients of LR image restoration loss $\mathcal{L}_\textit{RE}^p$ and $\frac{\partial \log p_{\textit{MCKA}}(\bm{\phi}_{\bm{k}}^{\textit{old}}|\bm{k}_{g})}{\partial \bm{\phi}_{\bm{k}}^{\textit{old}}}$ denotes the gradients of MSE loss $\mathcal{L}^l_\textit{MC}$. 
At each alternating step, the parameters of the kernel estimator are optimized with respect to the $\mathcal{L}_\textit{RE}^p$ and $\mathcal{L}^l_\textit{MC}$ in MLAO and MCKA phases, respectively. Specifically, the third term essentially plays the role of $\zeta_{\bm{n}}$ in vanilla Langevin dynamics (\ref{eq:langevin}), as the randomly sampled kernels $\bm{k}_{g}$ can be regarded as a "noise" in the view of network training samples. 
Besides, the sampling process follows MCMC simulation that also iteratively updates the sampling distribution with respect to the HR image restoration posterior. 
In this stage, we claim that this is a novel Langevin dynamics framework that is suitable for network-level optimization, as the conventional Langevin dynamics merely works on a traditional model-based optimization paradigm. This also provides theoretical support for rationalizing the incorporation of the MCKA phase.
}

\textcolor{black}{It is also noteworthy that the proposed network-level Langevin dynamics optimization framework is intrinsically different to the traditional Langevin dynamics, as such applied in BSRDM \cite{Yue2022blind}. In BSRDM, the input noise of DIP network is iteratively optimized via gradient-based algorithm and an additive white Gaussian noise is applied to prevent trapping into local mode. Differently, in the proposed MLMC, we proposed a novel network-level Langevin dynamics optimization strategy that is designed for network parameters optimization. Specifically, randomly sampled kernels are employed as the additive noises in this network-level Langevin dynamics optimization to play the role of preventing converging to a bad local optimum of the kernel estimator. Recalling the depiction in Fig. \ref{fig:MCKA}, the MCMC simulations on random Gaussian distributions will converge to a stable equilibrium, therefore, $\zeta$ is an organized random sample instead of total randomness. In this way, the proposed MCMC simulations on random Gaussian distributions bring a profitable disturbance towards the optimization process, as well as providing a rational kernel prior.}

\textcolor{black}{
\subsection{Pipeline and Analysis}\label{sec:Pipeline of the MLMC Approach}
The overall solution procedure of the proposed MLMC is given in Algorithm \ref{alg1:MLMC overall}, and the details of MCKA and MLAO loops are illustrated in Algorithm \ref{alg2:MCKA} and \ref{alg3:MLAO}, respectively.
}
At this stage, it is clear that the alternating framework between MCKA and MLAO is indispensable. On the one hand, at each iteration in Algorithm \ref{alg1:MLMC overall}, the MCKA phase learns a rational kernel prior from MCMC simulations, which provides external knowledge of the sampled kernel distributions. This contributes to a rational initialization of the $\bm{\phi}_{\bm{k},\textit{ML}}$ in MLAO stage instead of initializing from randomness. On the other hand, in the view of the total optimization on $\bm{\phi}_{\bm{k}}$, the MCKA phase provides organized randomness with the optimization based on LR observation in the MLAO phase, formulating the aforementioned network-level Langevin dynamics.

We note that the whole MLMC is processed in an unsupervised inference and the whole algorithm is applied in a plug-and-play fashion without any training in advance. This significantly improves the flexibility and generalization capacity of the MLMC.  In simulations, we show that the proposed MLMC is able to achieve strong generalization ability and flexibility towards out-of-distribution blur kernels, different noise scenarios and non-Gaussian kernels in Section \ref{sec:generalization results}, as well as realizing superior performance compared to the state-of-the-arts in Section \ref{sec:Experimental Results}. The hyper-parameters involved in the two summations corresponding to the number of sampling times in one MCMC simulation, denoted by $T$, and the meta-learning interval, denoted by $P$, will be discussed in Section \ref{sec:setup}.

\section{Experimental Results}\label{sec:Experimental Results}

\subsection{Experimental Setup} \label{sec:setup}

\begin{figure}[t]
\setlength{\abovecaptionskip}{0.1cm}
\setlength{\belowcaptionskip}{-0.4cm}
  \centering
  \includegraphics[width=0.9\linewidth]{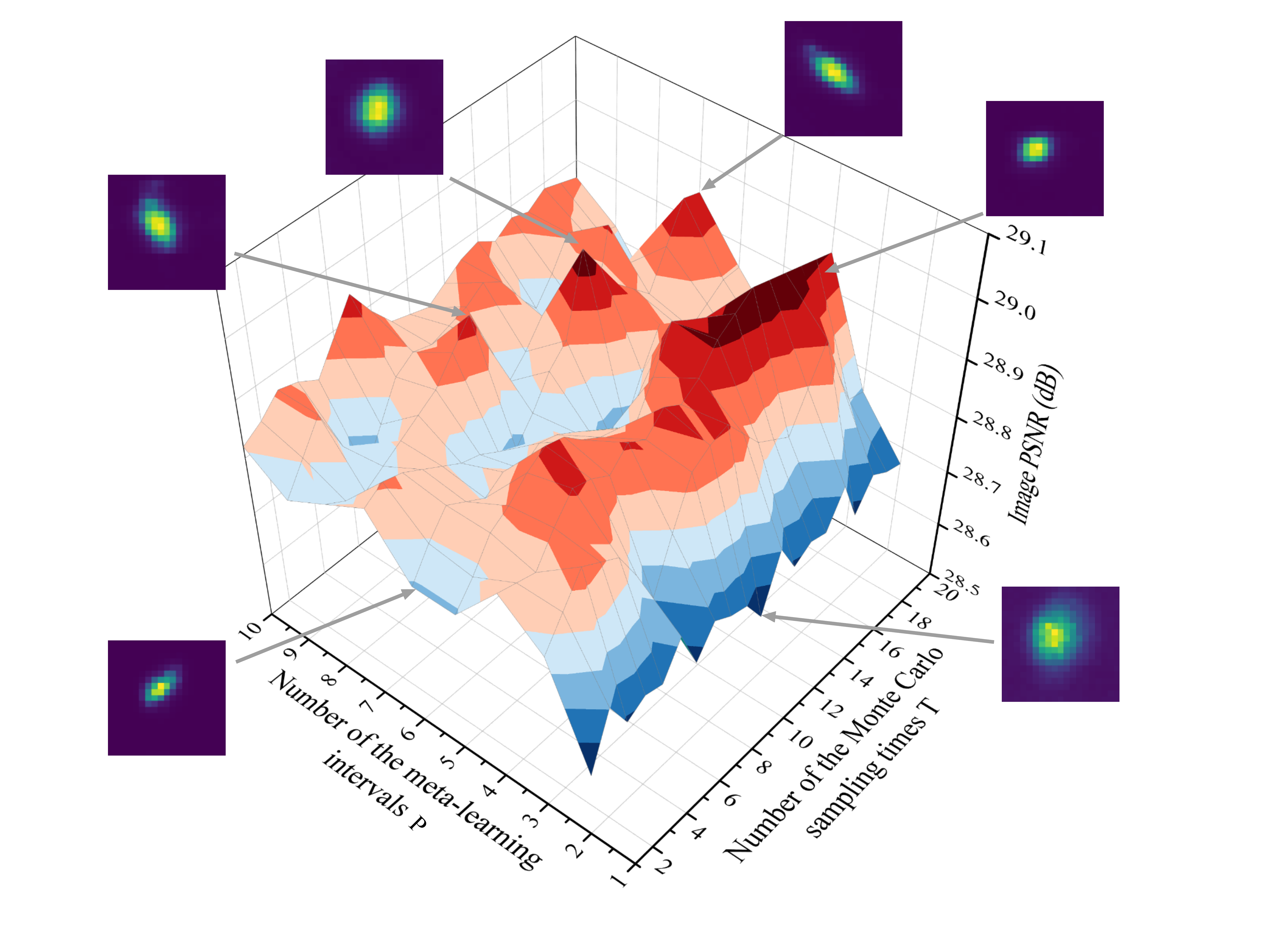}\\
  \caption{Image PSNR performance of the proposed MLMC method on Set5 with different hyper-parameter combinations: the number of the Monte Carlo sampling times $T$ and the number of meta-learning optimization intervals $P$.  
  }\label{fig:Parameters1}
\end{figure}

\noindent
\textbf{Data Preparation.} 
Following the widely adopted kernel assumption \cite{riegler2015conditioned, shao2015simple, shocher2018zero, soh2020meta, liang2021flow, Yue2022blind}, we conduct most of the experiments on anisotropic Gaussian kernels and a few on non-Gaussian kernels (motion kernels generated by \cite{kupyn2018deblurgan}).  The kernel sizes are set to $(4s+3)\times(4s+3)$, the Gaussian width ranges are set to $[0.175s, 2.5s]$, and the rotation angle range $\rho$ is set to $[0,\pi]$, for a scale factor  $s \in {2, 3, 4} $, respectively.
We synthesize LR images with random kernels with respect to Eq. (\ref{eq:degradation model}) for testing data based on four popular public benchmark datasets, including Set5 \cite{bevilacqua2012low}, Set14 \cite{zeyde2010single}, BSD100 \cite{martin2001database}, Urban100 \cite{huang2015single} and a real-world dataset, RealSRSet \cite{li2020efficient}.
We compare these kernels in terms of the peak signal to noise ratio (PSNR), and compare HR images in terms of PSNR and structural similarity (SSIM) \cite{wang2004image}.

\begin{table*}[t!] 
\setlength{\abovecaptionskip}{-0cm}
\setlength{\belowcaptionskip}{-0.2cm}
\caption{\textcolor{black}{Average PSNR/SSIM of different unsupervised methods (the model is optimized on only one LR image) on public datasets that are synthesized by the random Gaussian kernels with $s=2,3,4$. The best results are emphasized with \textbf{bold}. 
}
}
\begin{center} \label{table:images PSNR unsuperbised}
\small
\renewcommand{\arraystretch}{1}
\begin{tabular}{ |m{3.4cm}<{\centering} |m{0.9cm}<{\centering} |m{2.2cm}<{\centering} |m{2.2cm}<{\centering} |m{2.2cm}<{\centering} |m{2.2cm}<{\centering}|}
\hline
\text{Method}                    &Scale         &Set5            &Set14                    &BSD100                     &Urban100 \\
\hline
\end{tabular}

\vspace{2.5pt}

\begin{tabular}{ |m{3.4cm}<{\centering} |m{0.9cm}<{\centering} |m{2.2cm}<{\centering} |m{2.2cm}<{\centering} |m{2.2cm}<{\centering} |m{2.2cm}<{\centering}|}
\hline
    Bicubic         &$\times2$     & \color{black}{$26.58/0.8010$}      & \color{black}{$24.85/0.6939$}            & \color{black}{$25.19/0.6633$}             & \color{black}{$22.35/0.6503$}\\
    PAM                   &$\times2$     & $17.48/0.3990$      & $17.52/0.3948$            & $17.91/0.3826$             & $17.15/0.3763$\\
    DIP                          &$\times2$     & {$26.82/0.7518$}      & {$25.40/0.6868$}            & {$24.71/0.6508$}             & {$23.29/0.6749$}\\
    
    KernelGAN+ZSSR      &$\times2$     & {$26.45/0.7694$}      & {$24.64/0.6658$}           & {$23.96/0.6461$}             & {$22.02/0.6263$}\\
    Double-DIP                  &$\times2$     & $24.71/0.6423$      & $22.21/0.5626$            & $23.31/0.5681$             & $21.03/0.5701$\\
    FKP-DIP                    &$\times2$     & ${30.16/0.8637}$      & ${27.06/0.7421}$            & ${26.72/0.7089}$             & ${24.33/0.7069}$\\
    BSRDM                     &$\times2$     & ${30.56/0.8616}$      & ${27.15/0.7435}$            & ${26.69/0.7109}$             & ${24.52/0.7115}$\\
  \textbf{MLMC} (Ours)      &$\times2$     & $\bm{31.61/0.8836}$      & $\bm{28.52/0.7900}$            & $\bm{28.11/0.7751}$             & $\bm{25.32/0.7627}$\\
  \hline
\end{tabular}

\vspace{2.5pt}

\begin{tabular}{ |m{3.4cm}<{\centering} |m{0.9cm}<{\centering} |m{2.2cm}<{\centering} |m{2.2cm}<{\centering} |m{2.2cm}<{\centering} |m{2.2cm}<{\centering}|}
\hline
    Bicubic       &$\times3$     & \color{black}{$23.38/0.6836$}      & \color{black}{$22.47/0.5884$}            & \color{black}{$23.17/0.5625$}             & \color{black}{$20.37/0.5378$}\\
    PAM               &$\times3$     & ${15.60/0.3596}$      & ${16.23/0.3701}$            & ${16.41/0.3684}$             & ${15.23/0.3612}$\\
    DIP                     &$\times3$     & {$28.14/0.7687$}      & {$25.19/0.6581$}            & {$25.25/0.6408$}             & {$23.22/0.6512$}\\
    KernelGAN+ZSSR        &$\times3$     & ${25.57/0.6429}$      & ${23.51/0.6216}$            & ${23.08/0.6019}$             & ${21.98/0.5864}$\\
    Double-DIP                &$\times3$     & $23.21/0.6535$      & $20.20/0.5071$            & $20.38/0.4499$             & $19.61/0.4993$\\
    FKP-DIP                   &$\times3$     & ${28.82/0.8202}$      & ${26.27/0.6922}$            & ${25.96/0.6660}$             & ${23.47/0.6588}$\\
    BSRDM                    &$\times3$     & ${28.84/0.8255}$      & ${25.63/0.6973}$            & ${25.88/0.6576}$             & ${23.68/0.6783}$\\
  \textbf{MLMC} (Ours)      &$\times3$     & $\bm{30.21/0.8547}$      & $\bm{27.05/0.7363}$            & $\bm{26.77/0.7076}$             & $\bm{23.96/0.7050}$\\
    \hline
\end{tabular}

\vspace{2.5pt}

\begin{tabular}{ |m{3.4cm}<{\centering} |m{0.9cm}<{\centering} |m{2.2cm}<{\centering} |m{2.2cm}<{\centering} |m{2.2cm}<{\centering} |m{2.2cm}<{\centering}|}
\hline
    Bicubic       &$\times4$     & \color{black}{$21.70/0.6198$}      & \color{black}{$20.86/0.5181$}            & \color{black}{$21.95/0.5097$}             & \color{black}{$19.13/0.4729$}\\
    PAM     &$\times4$     & ${15.13/0.3938}$      & ${15.93/0.3849}$            & ${16.31/0.3894}$             & ${15.06/0.3784}$\\
    DIP                      &$\times4$     & {$27.34/0.7465$}      & {$25.03/0.6371$}            & {$24.92/0.6030$}             & {$22.55/0.6128$}\\
    KernelGAN+ZSSR      &$\times4$     & ${24.46/0.6216}$      & $22.65/0.5414$           & $21.49/0.5229$            & $21.04/0.4979$\\
    Double-DIP                 &$\times4$     & $20.99/0.5578$      & $18.31/0.6129$            & $18.57/0.3815$             & $18.15/0.4491$\\
    FKP-DIP                   &$\times4$     & ${27.77/0.7914}$      & $24.21/0.6684$            & ${25.15/0.6354}$             & ${22.89}/0.6327$\\
    BSRDM                   &$\times4$     & ${27.81/0.8029}$      & $25.35/{0.6859}$            & ${25.61/0.6526}$             & $22.36/{0.6601}$\\
  \textbf{MLMC} (Ours)      &$\times4$     & $\bm{28.87/0.8129}$      & $\bm{26.23/0.6938}$            & $\bm{25.89/0.6534}$             & $\bm{23.83/0.6728}$\\
  \hline

\end{tabular}
\end{center} 
\vspace{-0.2cm}
\end{table*}

\begin{table*}[t!] 
\setlength{\abovecaptionskip}{-0cm}
\setlength{\belowcaptionskip}{-0.2cm}
\caption{\textcolor{black}{Average PSNR/SSIM of different deep-learning-based methods (the model is pre-trained on LR-HR paired dataset) on public datasets that are synthesized by the random Gaussian kernels with $s=4$. The best results are emphasized with \textbf{bold}.
}
}
\begin{center} \label{table:images PSNR DL-based}
\small
\renewcommand{\arraystretch}{1}
\begin{tabular}{ |m{3.4cm}<{\centering} |m{0.9cm}<{\centering} |m{2.2cm}<{\centering} |m{2.2cm}<{\centering} |m{2.2cm}<{\centering} |m{2.2cm}<{\centering}|}
\hline
\text{Method}                    &Scale         &Set5            &Set14                    &BSD100                     &Urban100 \\
\hline
\end{tabular}

\vspace{2.5pt}

\begin{tabular}{ |m{3.4cm}<{\centering} |m{0.9cm}<{\centering} |m{2.2cm}<{\centering} |m{2.2cm}<{\centering} |m{2.2cm}<{\centering} |m{2.2cm}<{\centering}|}
\hline
    BSRGAN                   &$\times4$     & $20.47/0.5958$      & $19.92/0.4938$            & $20.89/0.4814$             & $17.61/0.3585$\\
    RCAN                   &$\times4$     & {$22.01/0.6210$}      & {$20.65/0.5031$}            & {$21.99/0.4986$}             & {$19.15/0.4503$}\\
    KDSR                  &$\times4$     & $23.35/0.6380$      & $21.73/0.5102$            & $23.85/0.5643$             & $20.44/0.4879$\\
    UDKE               &$\times4$     & $27.01/0.8038$      & $24.28/0.6923$            & $24.77/0.6695$             & $21.54/0.6515$\\
    DASR   &$\times4$     & $27.37/0.7859$      & $25.43/0.6591$            & $25.11/0.6129$             & $22.88/0.6448$\\
    KXNet                 &$\times4$     & $27.15/0.8085$      & $24.86/0.6680$            & $24.42/0.6431$             & $21.78/0.6543$\\
    KULNet                &$\times4$     & $27.89/0.8163$      & $25.43/0.6856$            & $25.03/0.6630$             & $21.98/0.6562$\\
    \textbf{MLMC} (Ours)      &$\times4$     & ${28.87/0.8129}$      & ${26.23/0.6938}$            & ${25.89/0.6534}$             & ${23.83/0.6728}$\\
    MANet+USRNet                   &$\times4$     & $\textcolor{black}{29.87/0.8572}$      & $\textcolor{black}{26.62/0.7360}$            & $\textcolor{black}{26.06/0.7080}$             & $\textcolor{black}{23.93/0.6944}$\\
  \textbf{MLMC}+USRNet (Ours)      &$\times4$     & $\bm{30.31/0.8607}$      & $\bm{27.46/0.7518}$            & $\bm{26.59/0.7121}$             & $\bm{23.98/0.7093}$\\
  \hline

\end{tabular}
\end{center} 
\vspace{-0.2cm}
\end{table*}

\noindent
\textbf{Implementation Details.}
We adopt a two-layer shallow FCN with 1000 nodes at each layer as the kernel generator, and use a CNN-based image generator following the architecture in \cite{ulyanov2018deep}.
The Adam optimizer \cite{kingma2014adam} is applied to optimize the parameters of these two networks with learning rates $0.5$ and $0.005$ for FCN and CNN, respectively. 
\textcolor{black}{We compare the proposed MLMC approach with other two kinds of state-of-the-arts, including unsupervised methods (the model is optimized on only one LR image): PAM \cite{perrone2015clearer}, DIP \cite{ulyanov2018deep}, KernelGAN+ZSSR \cite{shocher2018zero}, Double-DIP \cite{ren2020neural}, 
FKP-DIP \cite{liang2021flow} and BSRDM \cite{Yue2022blind},
and deep-learning-based methods (the model is pre-trained on LR-HR paired dataset): 
RCAN \cite{zhang2018image}, DASR \cite{wang2021unsupervised}, BSRGAN \cite{zhang2021designing}, KDSR \cite{xia2022knowledge}, KULNet \cite{fang2022uncertainty}, UDKE \cite{zheng2022unfolded}, KXNet \cite{fu2022kxnet}.
We also use non-blind model USRNet \cite{zhang2020deep} to generate the final SR result based on the kernel estimation from MANet \cite{liang2021mutual} and our MLMC as two deep-learning-based methods: MANet+USRNet and MLMC+USRNet.
}
\textcolor{black}{The parameters of the proposed MLMC approach are randomly initialized from scratch and re-initialized for each test image, ensuring fairness when compared with the deep-learning-based methods. For those methods that are originally applied to bicubic cases (RCAN), we also re-trained their model on the synthesized images with blur kernels before testing.}

\noindent
\textbf{Hyper-parameter Tuning.}
As mentioned in Sec. \ref{sec:methodology} the proposed MLMC has the following hyper-parameters: the MLMC algorithm adopts $I=100$, $L=1$, $Q=5$ for all the simulations. The number of the Monte Carlo sampling times $T$ and the number of meta-learning optimization intervals $P$. We present the HR image PSNR for different hyper-parameter settings in Fig. \ref{fig:Parameters1}. It can be seen that the performance fluctuates around the equilibrium PSNR when the meta-learning scheme is executed (P $\geq$ 2), and the variation on $T$ has little effect on the performance. To balance the efficiency and effectiveness, we thus set $T=10$ and $P=5$ for all the experiments. 


\begin{figure*}[bpht]
\setlength{\abovecaptionskip}{-0cm}
\setlength{\belowcaptionskip}{-0.1cm}
  \centering
  \includegraphics[width=0.9\linewidth]{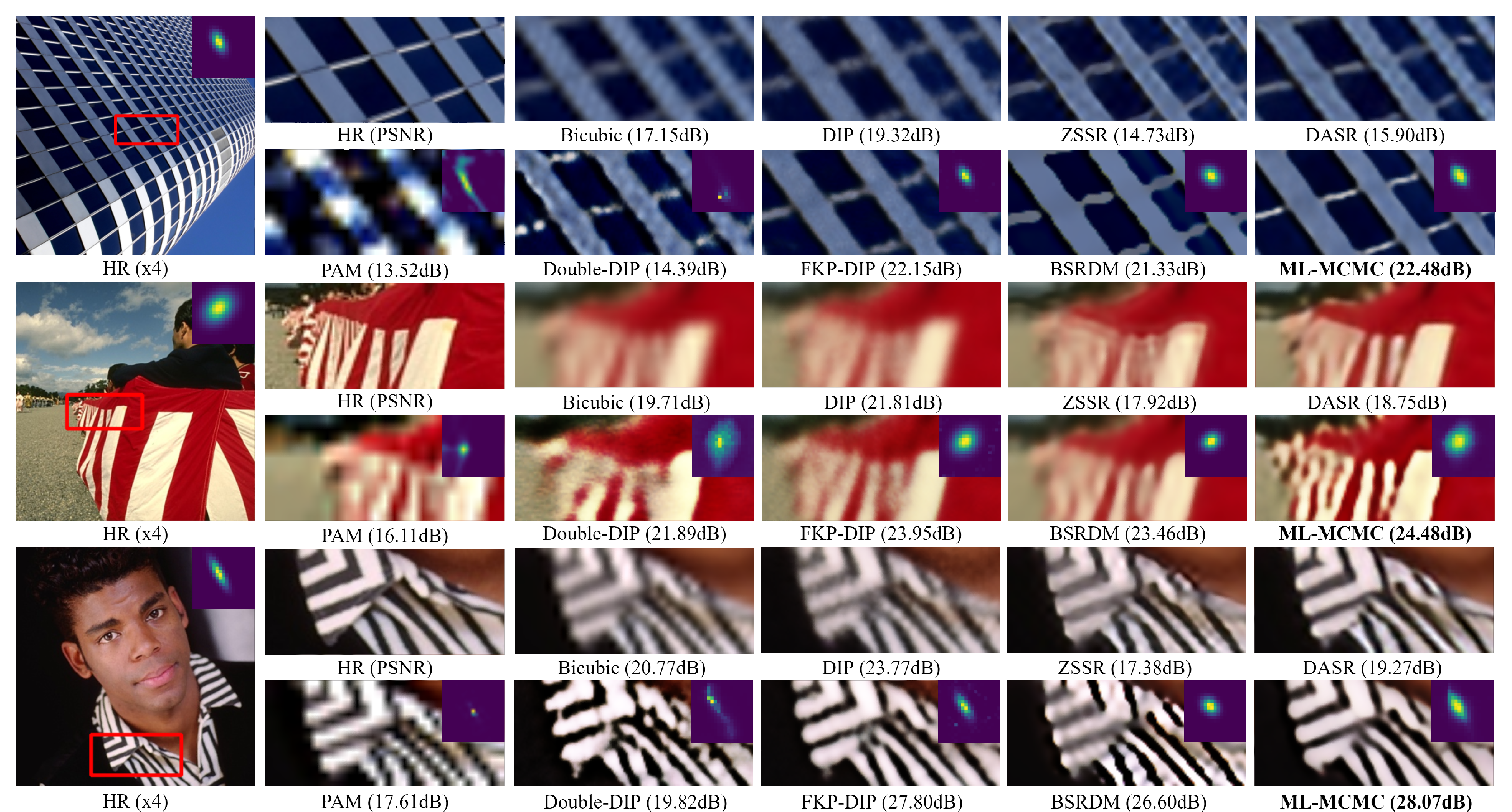}\\
  \caption{Visual results of different methods on public datasets for scale factor 4. Estimated/ground-truth kernels are shown on the top right.}\label{fig:visual results}
\end{figure*}

\begin{figure*}[bhtp]
\vspace{-0.1cm}
\setlength{\abovecaptionskip}{-0cm}
\setlength{\belowcaptionskip}{-0.1cm}
  \centering
  \includegraphics[width=0.9\linewidth]{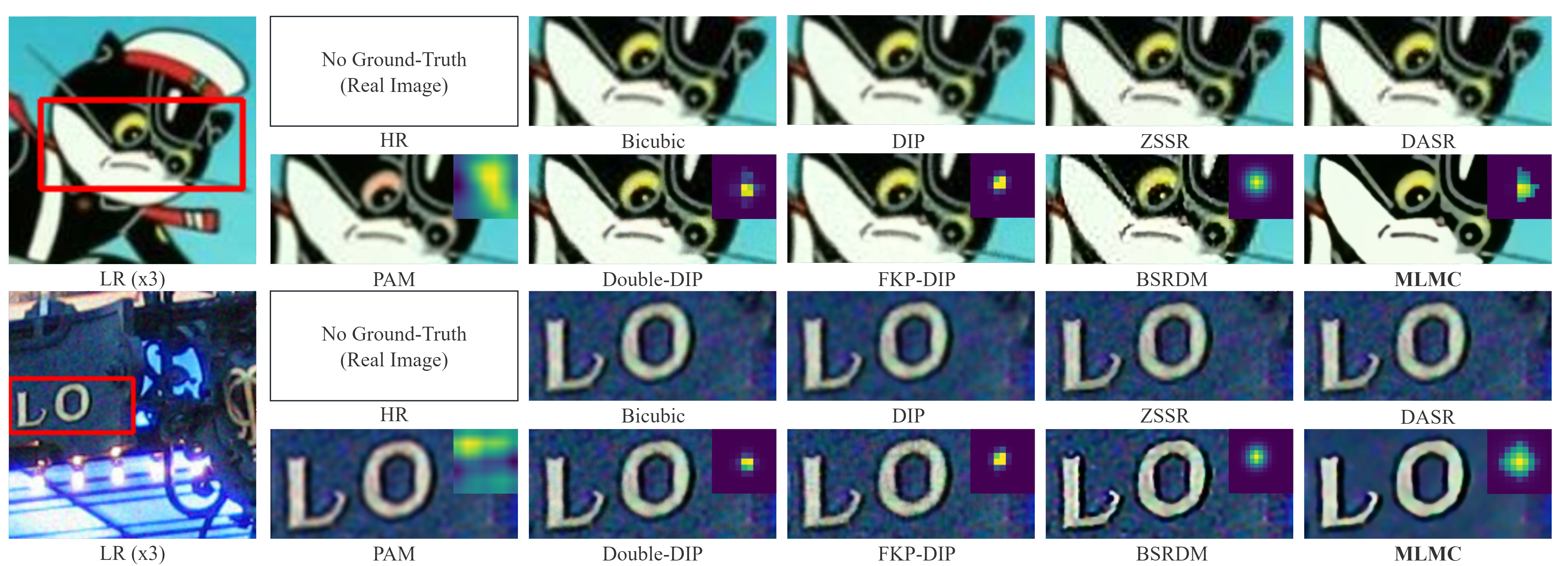}\\
  \caption{\textcolor{black}{Visual results of real world images on RealSRSet \cite{li2020efficient} for scale factor 3. Estimated kernels are shown on the top right.}}\label{fig:real visual results}
\end{figure*}

\begin{table}[t] 
\setlength{\abovecaptionskip}{-0cm}
\setlength{\belowcaptionskip}{-0cm}
\caption{Average PSNR/SSIM of image and PSNR of kernels on Set14. The best results are emphasized with \textbf{bold}.}
\begin{center} \label{table:kernel PSNR}
\renewcommand{\arraystretch}{1}
\begin{tabular}{|m{2.6cm}<{\centering} |m{0.6cm}<{\centering} |m{1.2cm}<{\centering} |m{2cm}<{\centering}|}
\hline
\text{Method}    &Scale  &Kernel PSNR      &Image PSNR/SSIM\\
\hline
\end{tabular}

\vspace{2.5pt}
\begin{tabular}{|m{2.6cm}<{\centering} |m{0.6cm}<{\centering} |m{1.2cm}<{\centering} |m{2cm}<{\centering}|}
\hline
\multicolumn{4}{|c |}{with Out-of-distribution Kernel}\\
\hline
  {Double-DIP}    &$\times2$     & ${46.22}$      & ${24.69/0.6374}$\\


  {FKP-DIP}       &$\times2$     & ${46.49}$      & ${26.13/0.7065}$\\

  {BSRDM}         &$\times2$     & ${42.53}$      & ${24.35/0.6728}$\\

  \textbf{MLMC} (Ours)      &$\times2$     & $\bm{52.32}$      & $\bm{26.63/0.7198}$\\

\hline
  {Double-DIP}    &$\times4$     & ${50.62}$      & ${20.51/0.4835}$\\


  {FKP-DIP}       &$\times4$     & ${54.46}$      & ${24.73/0.6344}$\\

  {BSRDM}         &$\times4$     & ${45.38}$      & ${21.57/0.5510}$\\

  \textbf{MLMC} (Ours)      &$\times4$     & $\bm{58.20}$      & $\bm{25.09/0.6474}$\\
\hline
\end{tabular}

\end{center} 
\vspace{-0.3cm}
\end{table}

\begin{table}[t] 
\setlength{\abovecaptionskip}{-0cm}
\setlength{\belowcaptionskip}{-0cm}
\caption{
\textcolor{black}{
Average PSNR/SSIM of image and PSNR of kernels on Set14. The best results are emphasized with \textbf{bold}.}
}
\begin{center} \label{table: image noise}
\renewcommand{\arraystretch}{1}
\begin{tabular}{|m{2.6cm}<{\centering} |m{0.6cm}<{\centering} |m{1.2cm}<{\centering} |m{2cm}<{\centering}|}
\hline
\text{Method}    &Scale  &Kernel PSNR      &Image PSNR/SSIM\\
\hline
\end{tabular}

\vspace{2.5pt}

\begin{tabular}{|m{2.6cm}<{\centering} |m{0.6cm}<{\centering} |m{1.2cm}<{\centering} |m{2cm}<{\centering}|}
\hline
\multicolumn{4}{|c |}{with Image Noise of $3.92\%$}\\
\hline
  {Double-DIP}             &$\times2$     & ${40.95}$      & ${23.93/0.6134}$\\

  {FKP-DIP}                &$\times2$     & ${44.38}$      & ${27.14/0.7228}$\\

  {BSRDM}                  &$\times2$     & ${43.96}$      & ${24.97/0.7069}$\\

  \textbf{MLMC} (Ours)     &$\times2$     & ${\bm{44.65}}$      & ${\bm{27.24/0.7345}}$\\
 
\hline
  {Double-DIP}             &$\times4$     & ${44.60}$      & ${18.42/0.3897}$\\

  {FKP-DIP}                &$\times4$     & ${47.96}$      & $\textcolor{black}{24.68/0.5760}$\\

  {BSRDM}                  &$\times4$     & ${48.73}$      & ${20.99/0.5387}$\\

  \textbf{MLMC} (Ours)     &$\times4$     & ${\bm{49.53}}$      & ${\bm{24.96/0.5951}}$\\
 
\hline
\end{tabular}

\vspace{2.5pt}

\begin{tabular}{|m{2.6cm}<{\centering} |m{0.6cm}<{\centering} |m{1.2cm}<{\centering} |m{2cm}<{\centering}|}
\hline
\multicolumn{4}{|c |}{with Image Noise of $7.84\%$}\\
\hline
  {Double-DIP}             &$\times2$     & ${40.23}$      & ${22.57/0.5376}$\\

  {FKP-DIP}                &$\times2$     & ${42.51}$      & ${26.21/0.6857}$\\

  {BSRDM}                  &$\times2$     & ${43.02}$      & ${24.59/0.6781}$\\

  \textbf{MLMC} (Ours)     &$\times2$     & ${\bm{45.84}}$      & ${\bm{26.37/0.6913}}$\\
 
\hline
  {Double-DIP}             &$\times4$     & ${42.88}$      & ${15.77/0.2628}$\\

  {FKP-DIP}                &$\times4$     & ${45.44}$      & ${24.06/0.5540}$\\

  {BSRDM}                  &$\times4$     & ${46.87}$      & ${20.82/0.5219}$\\

  \textbf{MLMC} (Ours)     &$\times4$     & ${\bm{47.20}}$      & ${\bm{24.53/0.5834}}$\\
 
\hline
\end{tabular}

\end{center} 
\vspace{-0.3cm}
\end{table}

\subsection{Comparisons with state-of-the-arts}

\noindent
\textbf{Quantitative Results of Unsupervised Methods.}
Quantitative evaluation results of unsupervised methods (the model is optimized on a single LR image) with scale factors from 2 to 4 are presented in Table \ref{table:images PSNR unsuperbised}, the best results are highlighted in \textbf{bold}. 
PAM is a classic model-based blind SISR approach, and it is difficult to get satisfactory performance when no prior is available.
Specifically, DIP and Double-DIP show less effectiveness due to the poor kernel estimation accuracy.
FKP-DIP and BSRDM achieve better results thanks to different pre-designed kernel priors. 
Meanwhile, the proposed MLMC approach achieves superior performance in all cases. \textcolor{black}{It is worth noting that the proposed MLMC significantly surpasses the counterpart unsupervised method BSRDM, especially for kernel estimation results. This recalls that BSRDM follows a gradient-based kernel estimation via explicit modeling on Gaussian distribution while MLMC formulates a network-level Langevin dynamic to learn from random kernel prior for better convergence performance.}

\noindent
\textcolor{black}{
\textbf{Quantitative Results of Deep-learning-based Methods.}
Quantitative evaluation results of DL-based methods (the model is pre-trained on LR-HR paired dataset) with scale factors 4 are presented in Table \ref{table:images PSNR DL-based}. The best and second best results are emphasized with \textbf{bold}.
Overall, the vanilla MLMC surpasses most of the DL-based supervised methods, showing comparable performance with KULNet and being next to MANet+USRNet. 
Besides, when the unsupervised MLMC is accompanied by the pre-trained image estimator USRNet, the proposed MLMC+USRnet is able to realize slightly better performance than the supervised MANet+USRNet. 
}

\noindent
\textbf{Visual Results.}
Fig. \ref{fig:visual results} presents the visually qualitative results from Set14 \cite{zeyde2010single}, BSD100 \cite{martin2001database} and Urban100 \cite{huang2015single}, while  Fig. \ref{fig:real visual results} further shows visual results from RealSRSet \cite{li2020efficient}. Apparently, the proposed MLMC obtains the most concise blur kernels as well as restored HR images, almost keeping consistence with the ground truth, while FKP-DIP and BSRDM show different levels of distortion on the estimated kernels, and Double-DIP fails to estimate reasonable blur kernels. Particularly, the real-world image test results demonstrate that all the approaches except MLMC estimate a Gaussian-like blur kernel, whereas MLMC tends to find a non-Gaussian mode. This also verifies that an adaptive and flexible kernel estimation principle is learned from the alternative Monte Carlo simulations, and hence, fitting the real-world application better. 

\subsection{Generalization to out-of-distribution kernels}\label{sec:generalization results}
To further evaluate the generalization-ability of the proposed MLMC approach, it is compared with Double-DIP, BSRDM, and FKP-DIP, which use different kernel priors in more challenging cases, including out-of-distribution kernels (unseen Gaussian blur kernels with larger width range $[0.35s, 5s]$ than assumption kernel distribution within width range $[0.175s, 2.5s]$).
As we can see from Table \ref{table:kernel PSNR}, the proposed MLMC significantly outperforms the competitors in the case with out-of-distribution kernel by large margins.
We note that FKP-DIP experiences performance drops in these challenge cases, due to the high dependency on pre-trained kernel priors, while MLMC still shows superior performance.
In this case, we conclude that MLMC has superior generalization-ability to arbitrary kernels and exhibits satisfactory robustness to unseen kernels and non-Gaussian kernels, even without specific kernel priors.

\subsection{\textcolor{black}{Generalization to motion kernels}}
\begin{figure}[t!]
  \centering
  \includegraphics[width=1\linewidth]{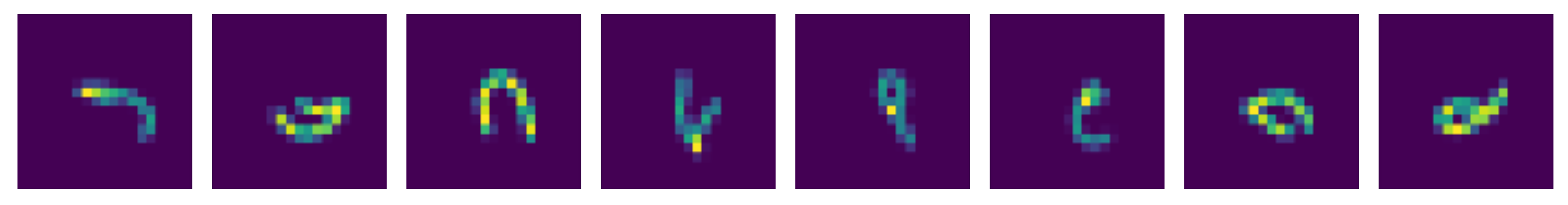}
  \vspace{-0.4cm}
  \caption{\textcolor{black}{The visualization of the random motion blur kernels.}
  }
  \vspace{-0.1cm}
  \label{fig:motion kernel visual}
\end{figure}

\textcolor{black}{The proposed MLMC can be directly expended on solving motion kernel tasks without re-training. The fundamental technique process keeps consistent with the Gaussian kernel case, since our MLMC only replaces the random kernel generation function following settings in \cite{kupyn2018deblurgan}. 
We visualize the generated random motion kernels in Fig. \ref{fig:motion kernel visual}. 
Table \ref{table:images PSNR motion} presents the quantitative evaluation results of different methods in the motion kernel scenario.
Compared with the Gaussian kernel scenario, all the methods show different degrees of performance drop, while the proposed MLMC shows significantly better performance than other alternatives.
This verifies the better generalization ability of our MLMC towards more complex degradation kernels.}

\begin{table}[t] 
\setlength{\abovecaptionskip}{-0cm}
\setlength{\belowcaptionskip}{-0cm}
\caption{\textcolor{black}{Average PSNR/SSIM of image and PSNR of kernels on Set14 and BSD100 in the motion kernel scenario. The best results are emphasized with \textbf{bold}.}
}
\begin{center} \label{table:images PSNR motion}
\renewcommand{\arraystretch}{1}
\begin{tabular}{|m{2.6cm}<{\centering} |m{0.6cm}<{\centering} |m{1.2cm}<{\centering} |m{2cm}<{\centering}|}
\hline
\text{Method}    &Scale  &Kernel PSNR      &Image PSNR/SSIM\\
\hline
\end{tabular}

\vspace{2.5pt}

\begin{tabular}{|m{2.6cm}<{\centering} |m{0.6cm}<{\centering} |m{1.2cm}<{\centering} |m{2cm}<{\centering}|}
\hline
\multicolumn{4}{|c |}{with Motion Kernel on Set14}\\
\hline
  {Double-DIP}             &$\times2$     & ${31.07}$      & ${22.80/0.6449}$\\

  {FKP-DIP}                &$\times2$     & ${33.97}$      & ${25.99/0.8157}$\\

  {BSRDM}                  &$\times2$     & ${30.31}$      & ${23.62/0.6447}$\\

  \textbf{MLMC} (Ours)     &$\times2$     & $\bm{34.44}$      & $\bm{28.48/0.8327}$\\
 
\hline
  {Double-DIP}             &$\times4$     & ${37.30}$      & ${20.96/0.4817}$\\

  {FKP-DIP}                &$\times4$     & ${38.07}$      & ${22.79/0.6278}$\\

  {BSRDM}                  &$\times4$     & ${36.27}$      & ${21.12/0.5491}$\\

  \textbf{MLMC} (Ours)     &$\times4$     & $\bm{39.44}$      & $\bm{25.22/0.6776}$\\
 
\hline
\end{tabular}

\vspace{2.5pt}

\begin{tabular}{|m{2.6cm}<{\centering} |m{0.6cm}<{\centering} |m{1.2cm}<{\centering} |m{2cm}<{\centering}|}
\hline
\multicolumn{4}{|c |}{with Motion Kernel on BSD100}\\
\hline
  {Double-DIP}             &$\times2$     & ${29.94}$      & ${22.31/0.6388}$\\

  {FKP-DIP}                &$\times2$     & ${32.49}$      & ${25.27/0.7784}$\\

  {BSRDM}                  &$\times2$     & ${30.63}$      & ${23.88/0.5999}$\\

  \textbf{MLMC} (Ours)     &$\times2$     & $\bm{33.93}$      & $\bm{28.10/0.8087}$\\
 
\hline
  {Double-DIP}             &$\times4$     & ${35.45}$      & ${18.69/0.3646}$\\

  {FKP-DIP}                &$\times4$     & ${37.06}$      & ${23.07/0.6264}$\\

  {BSRDM}                  &$\times4$     & ${36.65}$      & ${21.31/0.5543}$\\

  \textbf{MLMC} (Ours)     &$\times4$     & $\bm{40.08}$      & $\bm{25.58/0.6554}$\\
 
\hline
\end{tabular}
\end{center} 
\vspace{-0.3cm}
\end{table}

\subsection{Robustness to image noise}
We add image noise ( $3.92\%$ and $7.84\%$ of the maximum image pixel value) to the LR image after blurring and downsampling.
In Table \ref{table: image noise}, the proposed MLMC produces comparable result in all the cases, and shows good robustness to different levels of image noise.
Although MLMC has a modest performance drop when the image is corrupted by noise, it still surpasses all of the other comparative methods.
In this case, we argue that MLMC is able to handle a more complicated degradation model with better robustness.


\begin{figure*}[htbp]
\centering
\subfigure[``005" in Urban100 \cite{huang2015single}]{
\includegraphics[width=0.31\linewidth]{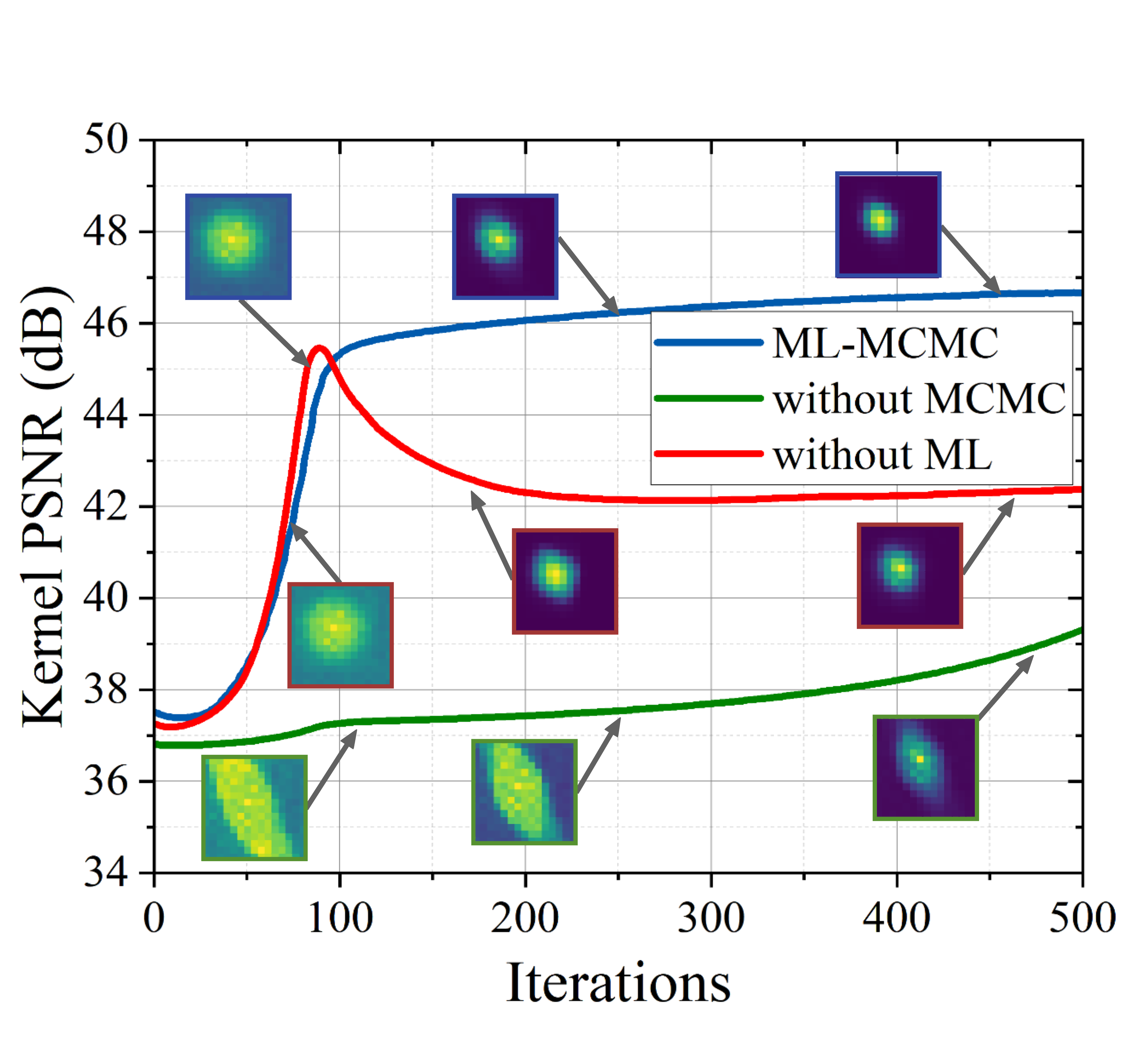}
}
\hspace*{\fill}
\subfigure[``067" in BSD100 \cite{martin2001database}]{
\includegraphics[width=0.31\linewidth]{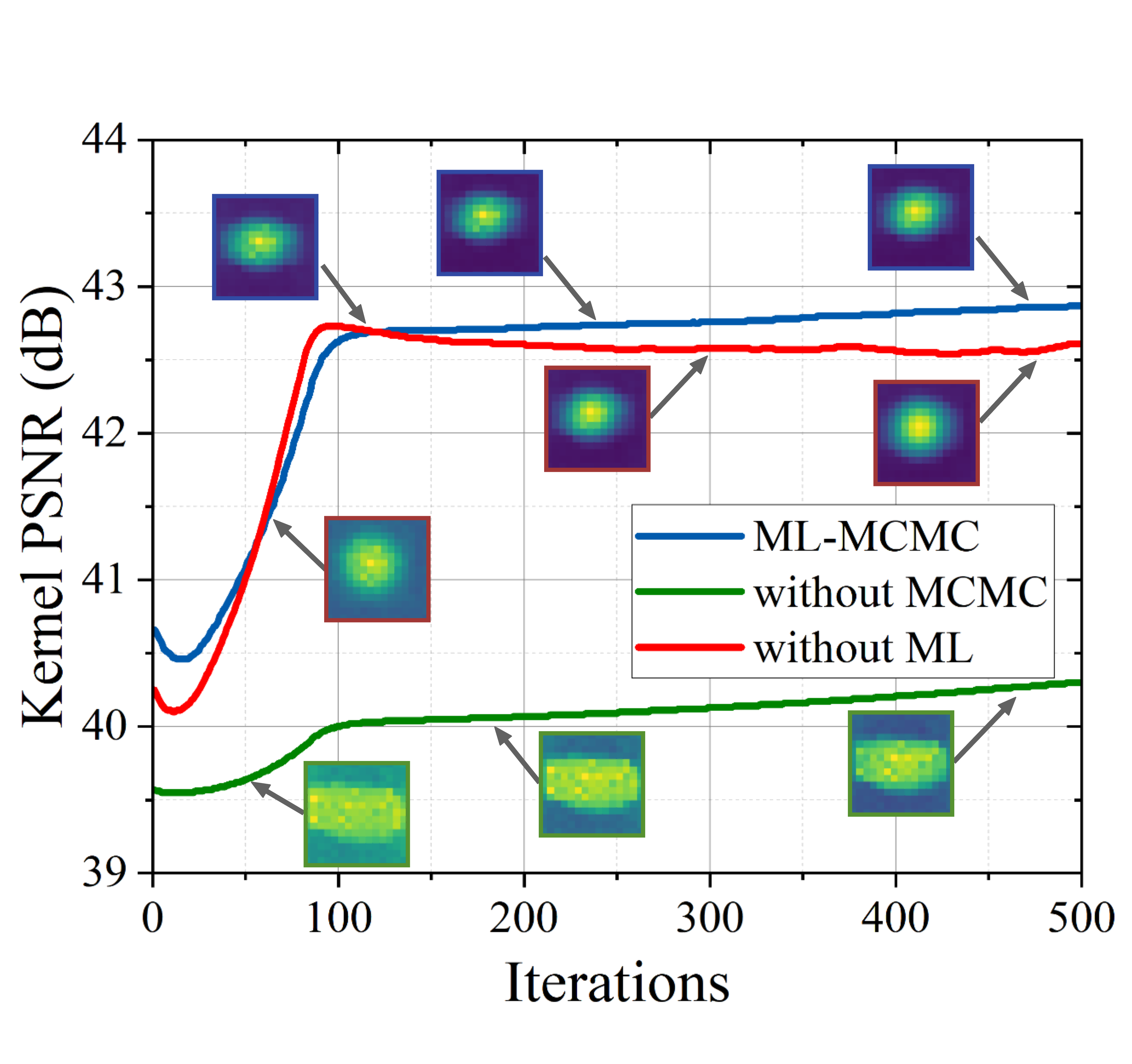}
}
\hspace*{\fill}
\subfigure[``071" in BSD100 \cite{martin2001database}]{
\includegraphics[width=0.31\linewidth]{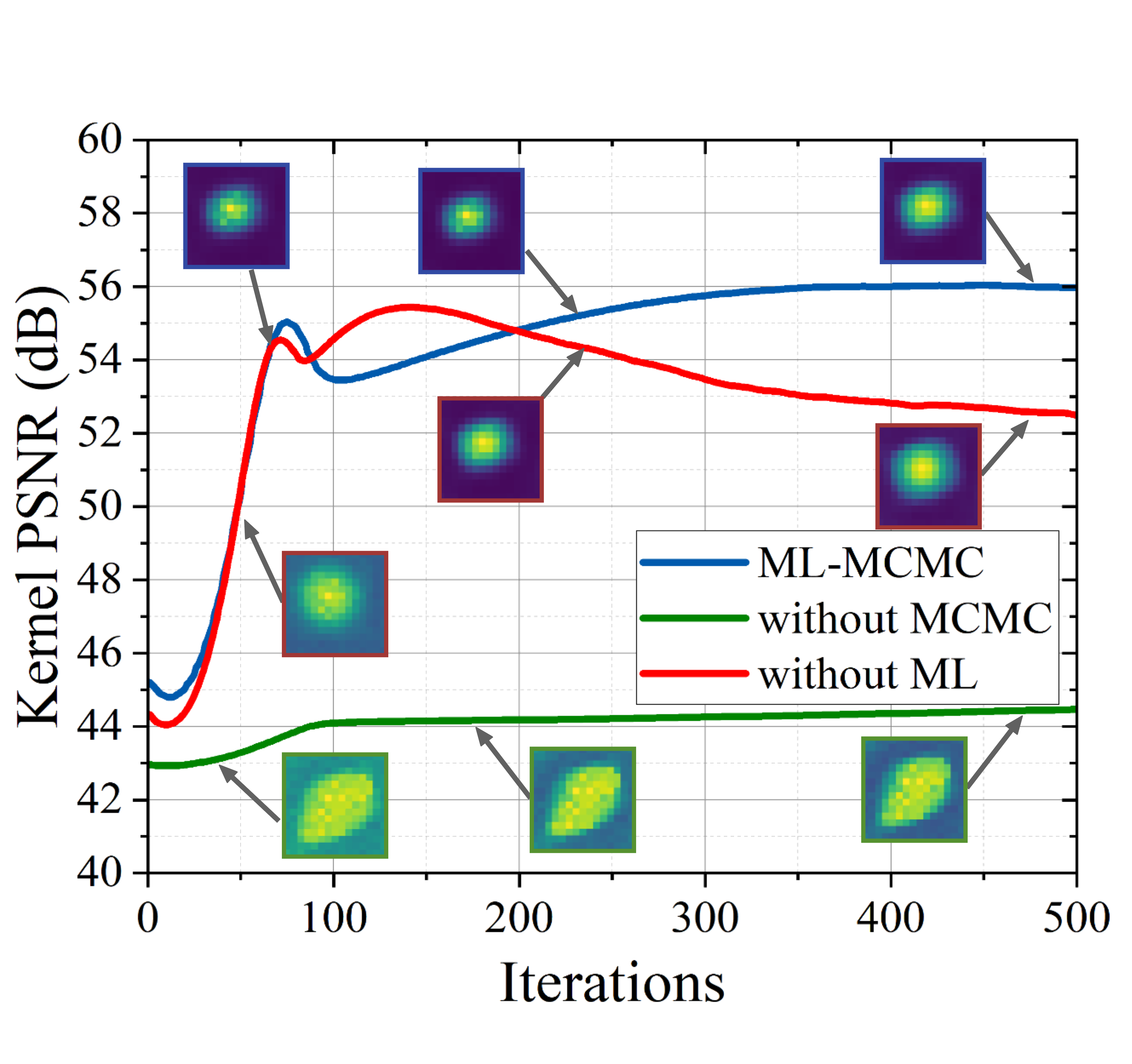}
}
\caption{The intermediate kernel results of MLMC, MLMC without Monte Carlo and MLMC without meta-learning over iterations on three test images.}
\label{fig:Ablations-3}
\end{figure*}

 \begin{table}[ht!] 
\setlength{\abovecaptionskip}{-0cm}
\setlength{\belowcaptionskip}{-0cm}
\caption{\textcolor{black}{Ablation results on Set14 and BSD100 with scale factor $2$ and $4$. The best results are emphasized with \textbf{bold}.}
}
\begin{center} \label{table:ablation Set14}
\footnotesize
\renewcommand{\arraystretch}{1}
\begin{tabular}{|m{2.9cm}<{\centering} |m{0.6cm}<{\centering} |m{1.3cm}<{\centering} m{2cm}<{\centering}|}
\hline
\text{Method}    &Scale  &Kernel PSNR      &Image PSNR/SSIM\\
\hline
\end{tabular}

\vspace{2.5pt}
\begin{tabular}{|m{2.9cm}<{\centering} |m{0.6cm}<{\centering} |m{1.3cm}<{\centering} m{2cm}<{\centering}|}
\hline
\multicolumn{4}{|c |}{with Gaussian Kernel on Set14}\\
\hline

  without Monte Carlo     &$\times2$     & {${41.30}$}      & {${24.23/0.6318}$}\\

  without Meta-learning   &$\times2$     & ${42.54}$      & ${26.94/0.7353}$\\

  \textbf{MLMC} (Ours)    &$\times2$     & $\bm{45.01}$      & $\bm{28.52/0.7900}$\\
\hline

  without Monte Carlo     &$\times4$     & {${44.51}$}      & {${19.83/0.4999}$}\\

  without Meta-learning   &$\times4$     & ${51.39}$      & ${25.01/0.6869}$\\

  \textbf{MLMC} (Ours) &$\times4$     & $\bm{55.98}$      & $\bm{26.23/0.6938}$\\
\hline
\end{tabular}

\vspace{2.5pt}
\begin{tabular}{|m{2.9cm}<{\centering} |m{0.6cm}<{\centering} |m{1.3cm}<{\centering} m{2cm}<{\centering}|}
\hline
\multicolumn{4}{|c |}{with Gaussian Kernel on BSD100}\\
\hline

  without Monte Carlo     &$\times2$     & {${37.58}$}      & {${22.14/0.5340}$}\\

  without Meta-learning   &$\times2$     & ${40.43}$      & ${25.76/0.7020}$\\

  \textbf{MLMC} (Ours)    &$\times2$     & $\bm{47.84}$      & $\bm{28.11/0.7751}$\\
\hline

  without Monte Carlo     &$\times4$     & {${49.17}$}      & {${20.56/0.4782}$}\\

  without Meta-learning   &$\times4$     & ${51.93}$      & ${24.85/0.6210}$\\

  \textbf{MLMC} (Ours)   &$\times4$     & $\bm{53.16}$      & $\bm{25.89/0.6534}$\\
\hline
\end{tabular}
\end{center} 
\vspace{-0.4cm}
\end{table}

\subsection{Ablation Studies}\label{sec-ablation}
In Fig. \ref{fig:Ablations-3}, intermediate kernel results of three ablation experiments are depicted to highlight the effectiveness of the introduced Monte Carlo kernel approximation and meta-learning SISR stages, respectively. 
It can be seen that the estimated kernel without Monte Carlo simulations has non-negligible drop in PSNR as well as significant distortion, compared to MLMC. Meanwhile, the absence of the meta-learning scheme leads to a significant visible-deviation during the optimization, indicating a different and worse local optima compared to the one MLMC reached. 

\begin{figure}[t]
\vspace{-0.1cm}
\setlength{\abovecaptionskip}{-0cm}
\setlength{\belowcaptionskip}{-0.1cm}
  \centering
  \includegraphics[width=1\linewidth]{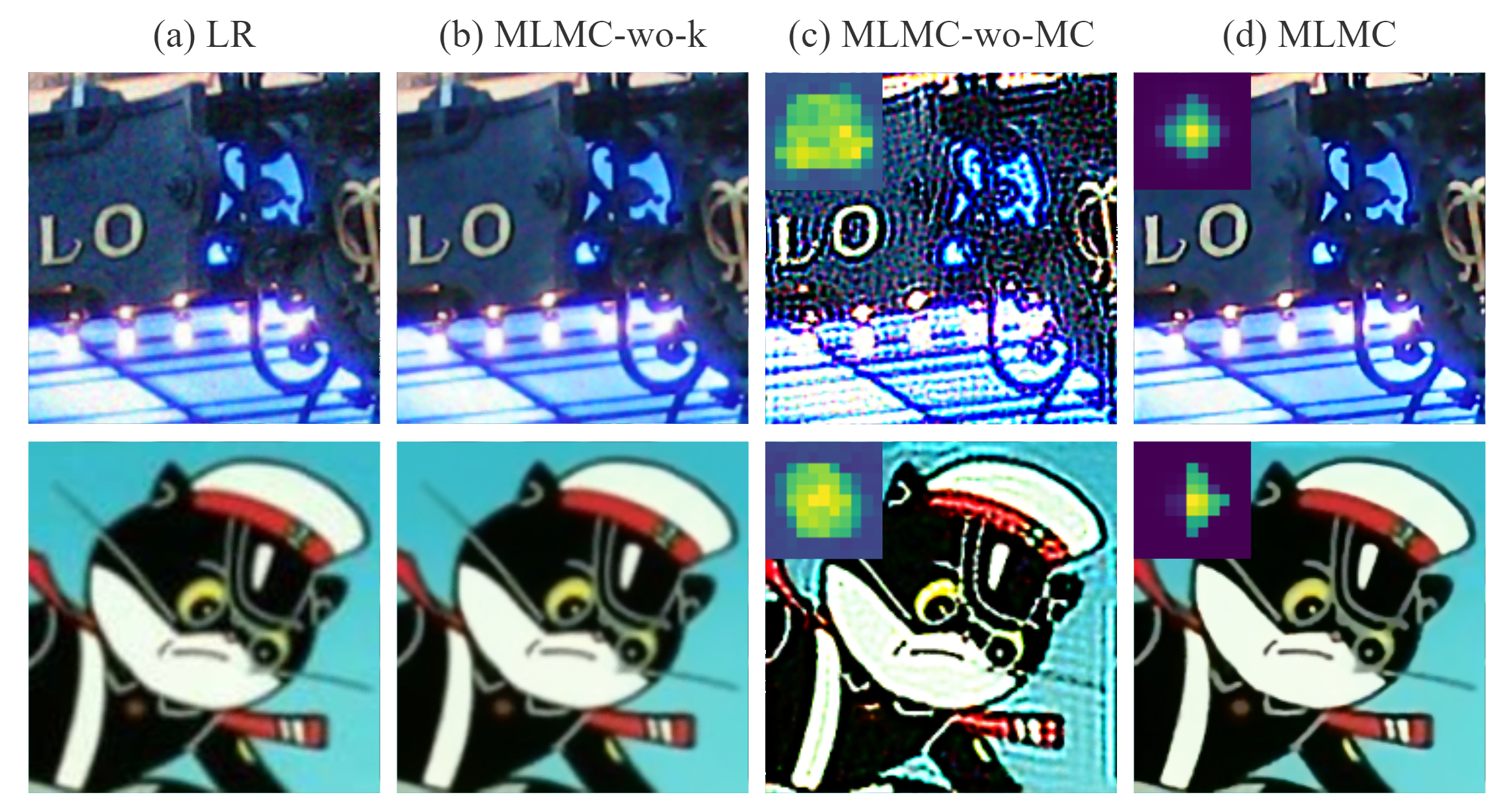}\\
  \caption{\textcolor{black}{
  Three ablation cases: (a) LR image, (b) MLMC-wo-k (MLMC without the kernel estimation process), (c) MLMC-wo-MC (MLMC without the Monte-Carlo sampling process), and (d) original MLMC.
  } }\label{fig: ablation of real visual results}
\end{figure}

\textcolor{black}{
In Table \ref{table:ablation Set14}, we further present the average results of the ablation experiments of the proposed Monte Carlo kernel approximation and meta-learning SISR stages under two different scale factors: $2$ and $4$, two datasets: Set14 \cite{zeyde2010single} and BSD100 \cite{martin2001database}.
We can see that the proposed MLMC achieves the best results for all the cases. 
As we analyze in the Section \ref{sec:kernel two stage} and Section \ref{sec:meta-learning} that 
on the one hand, without Monte Carlo simulations, the estimated kernels are not regular, which leads to bad kernel estimation and image restoration performance.
On the other hand, without the meta-learning SISR scheme, the estimated kernels are easily converged to the local optimum, which results in unsatisfactory image restoration performance. For kernel estimation, we also evaluate the necessity of the MCMC simulation towards the kernel prior learning. Fig. \ref{fig: ablation of real visual results} shows the visualization of ablation on real-world images from three cases. It is clear that without MCMC simulation, the kernel estimation no longer provides performance improvements on HR image restoration and the estimated kernels are distorted. Meanwhile, we can see that the MLMC tends to estimate non-Gaussian kernels via combinations of Gaussian sampling results to fit the real-world degradations.
}

\textcolor{black}{In table \ref{R3C4 table: ablation Set5}, we validate the effects of different regularizers in Eq. (\ref{eq:image noise loss}). The hyper-Laplacian, Tikhnonv, and TV regularizers are compared in two cases with different levels of noise. In both cases, hyper-Laplacian achieves the best performance, indicating the robustness towards noise level variation.}

 \begin{table}[ht!] 
\setlength{\abovecaptionskip}{-0cm}
\setlength{\belowcaptionskip}{-0cm}
\caption{\textcolor{black}{Ablation results of regularizers in Eq. \eqref{eq:image noise loss} of our MLMC on Set5 with scale factor $4$. R1: Hyper-Laplacian, R2: Tikhnonv, R3: TV. The best results are emphasized with \textbf{bold}.}}
\vspace{-0.3cm}
\begin{center} \label{R3C4 table: ablation Set5}
\renewcommand{\arraystretch}{1}
\small

\vspace{2.5pt}
\begin{tabular}{|m{0.8cm}<{\centering} |m{0.8cm}<{\centering} |m{0.8cm}<{\centering} |m{1.4cm}<{\centering}|m{2.4cm}<{\centering}|}
\hline
\text{R1} & \text{R2}  & \text{R3} &Kernel PSNR   &Image PSNR/SSIM\\
\hline
\end{tabular}

\vspace{2.5pt}
\begin{tabular}{|m{0.8cm}<{\centering} |m{0.8cm}<{\centering} |m{0.8cm}<{\centering} |m{1.4cm}<{\centering}|m{2.4cm}<{\centering}|}
\hline
\multicolumn{5}{|c |}{with Image Noise of 3.92\%, Scale $\times 4$}\\
\hline

    \XSolidBrush  & \XSolidBrush  & \XSolidBrush       & ${49.38}$   & ${26.25/0.6448}$\\

    \Checkmark    & \XSolidBrush  & \XSolidBrush       & $\bm{49.40}$   & $\bm{27.65/0.7707}$\\     
    
    \XSolidBrush  & \Checkmark    & \XSolidBrush     & ${48.87}$   & ${26.71/0.6824}$\\
    
    \XSolidBrush  & \XSolidBrush  & \Checkmark       & ${49.04}$   & ${26.82/0.6967}$\\

\hline
\end{tabular}

\vspace{2.5pt}
\begin{tabular}{|m{0.8cm}<{\centering} |m{0.8cm}<{\centering} |m{0.8cm}<{\centering} |m{1.4cm}<{\centering}|m{2.4cm}<{\centering}|}
\hline
\multicolumn{5}{|c |}{with Image Noise of 7.84\%, Scale $\times 4$}\\
\hline

    \XSolidBrush  & \XSolidBrush  & \XSolidBrush     & ${45.57}$   & ${24.07/0.5163}$\\

    \Checkmark    & \XSolidBrush  & \XSolidBrush     & $\bm{46.05}$   & $\bm{25.57/0.6493}$\\     
    
    \XSolidBrush  & \Checkmark    & \XSolidBrush     & ${45.40}$   & ${24.59/0.5576}$\\

    \XSolidBrush  & \XSolidBrush  & \Checkmark       & ${45.53}$   & ${24.63/0.5598}$\\

\hline
\end{tabular}
\end{center} 
\vspace{-0.4cm}
\end{table}

\begin{table}[ht!] 
\vspace{0cm}
\setlength{\abovecaptionskip}{0cm}
\setlength{\belowcaptionskip}{-0cm}
\caption{Comparison of different methods from tested LR image size of $256 \times 256$ with scale factor $s=2,3,4$. Computational requirements are model size (K) and runtime (s).}
\vspace{-0.2cm}
\begin{center} \label{table:computational complexity1}
\renewcommand{\arraystretch}{1}
\begin{tabular}{|m{2.5cm}<{\centering} |m{0.6cm}<{\centering} |m{1.1cm}<{\centering} m{1.9cm}<{\centering} m{0.7cm}<{\centering}|}
\hline
\text{Method}  &Scale  &Pre-training  &Model Size  & Time\\
\hline
\end{tabular}

\vspace{2.5pt}

\begin{tabular}{|m{2.5cm}<{\centering} |m{0.6cm}<{\centering} |m{1.1cm}<{\centering} m{1.9cm}<{\centering} m{0.7cm}<{\centering}|}
\hline
    ZSSR  \cite{shocher2018zero}      &$\times2$  &\XSolidBrush     & $225K$       & $56s$\\
    Double-DIP   \cite{ren2020neural} &$\times2$       &\XSolidBrush     & $2359K+641K$       & $47s$\\
    FKP-DIP  \cite{liang2021flow}     &$\times2$       &$6$ hours       & $2359K+195K$       & $45s$\\
    BSRDM    \cite{Yue2022blind}      &$\times2$      &\XSolidBrush     & $766K$             & $37s$\\
  \textbf{MLMC} (Ours)             &$\times2$          &\XSolidBrush     & $2359K+562K$       & $41s$\\
  \hline

    ZSSR  \cite{shocher2018zero}      &$\times4$  &\XSolidBrush     & $225K$       & $235s$\\
    Double-DIP   \cite{ren2020neural} &$\times4$       &\XSolidBrush     & $2359K+641K$       & $239s$\\
    FKP-DIP  \cite{liang2021flow}     &$\times4$       &$6$ hours        & $2359K+143K$       & $232s$\\
    BSRDM    \cite{Yue2022blind}      &$\times4$      &\XSolidBrush     & $766K$       & $190s$\\
  \textbf{MLMC} (Ours)             &$\times4$          &\XSolidBrush     & $2359K+562K$       & $215s$\\
  \hline
\end{tabular}
\end{center} 
\vspace{-0.3cm}
\end{table}

\subsection{Model Size, Runtime and Memory Usage}
Table \ref{table:computational complexity1} compares the results on model size (number of parameters), runtime and pre-training requirements of the four approaches. All simulations are accelerated by GeForce RTX 3090 GPU. The input LR images are with size $256\times256$ and scale factors $s=2,3,4$. It can be seen that the MLMC approach has a similar model size and enjoys competitive runtimes compared to the latest learning-based approaches. Meanwhile, pre-training-based approaches, such as FKP-DIP, which typically requests 5-6 hours for pre-training. \textcolor{black}{The memory usage of our MLMC on a GeForce RTX 3090 GPU for generating an HR image of size $1024\times1024$ is about $11$GB memory, which is close to the Double-DIP ($11.2$GB) and DIP-FKP ($10.6$GB). 
We note that the plug-and-play fashion and the better flexibility towards unknown degradations allow significant merits in dealing with blind SR tasks with real-world scenarios, especially those scenarios without high-quality training data and with complex blurring, such as space high-speed targets (e.g., satellites, aircraft) and medical images (e.g., beating heart).}

\section{Conclusion}
In this paper, we have proposed a new learning-based blind SISR method, which combines MCMC simulations and meta-learning-based optimization to achieve superior kernel estimation. Most strikingly, the proposed approach does not require any supervised pre-training or parametric priors.
\textcolor{black}{In future work, we will investigate two main directions for better practicality of the proposed MLMC methods, including i) the expansion to other degradation models, such as compression artifacts, deraining, and shadow-removal, to improve the generalization ability; and ii) the application with more advanced pre-trained SR models, for example, USRNet and diffusion models, to play the role of kernel prior learning module for performance improvements.}
We believe that the concept introduced here, in particular, learning from randomness to provide priors and the meta-learning-based non-convex optimization algorithm, will lead to a new direction of solving blind image restoration tasks to achieve superior performance with limited computational complexity.

\ifCLASSOPTIONcaptionsoff
  \newpage
\fi

\bibliographystyle{IEEEtran}
\bibliography{reference}

\begin{thebibliography}{10}
\providecommand{\url}[1]{#1}
\csname url@samestyle\endcsname
\providecommand{\newblock}{\relax}
\providecommand{\bibinfo}[2]{#2}
\providecommand{\BIBentrySTDinterwordspacing}{\spaceskip=0pt\relax}
\providecommand{\BIBentryALTinterwordstretchfactor}{4}
\providecommand{\BIBentryALTinterwordspacing}{\spaceskip=\fontdimen2\font plus
\BIBentryALTinterwordstretchfactor\fontdimen3\font minus \fontdimen4\font\relax}
\providecommand{\BIBforeignlanguage}[2]{{%
\expandafter\ifx\csname l@#1\endcsname\relax
\typeout{** WARNING: IEEEtran.bst: No hyphenation pattern has been}%
\typeout{** loaded for the language `#1'. Using the pattern for}%
\typeout{** the default language instead.}%
\else
\language=\csname l@#1\endcsname
\fi
#2}}
\providecommand{\BIBdecl}{\relax}
\BIBdecl

\bibitem{perrone2015clearer}
D.~Perrone and P.~Favaro, ``A clearer picture of total variation blind deconvolution,'' \emph{IEEE transactions on pattern analysis and machine intelligence}, vol.~38, no.~6, pp. 1041--1055, 2015.

\bibitem{ulyanov2018deep}
D.~Ulyanov, A.~Vedaldi, and V.~Lempitsky, ``Deep image prior,'' in \emph{Proceedings of the IEEE conference on computer vision and pattern recognition}, 2018, pp. 9446--9454.

\bibitem{bell2019blind}
S.~Bell-Kligler, A.~Shocher, and M.~Irani, ``Blind super-resolution kernel estimation using an internal-gan,'' \emph{Advances in Neural Information Processing Systems}, vol.~32, 2019.

\bibitem{ren2020neural}
D.~Ren, K.~Zhang, Q.~Wang, Q.~Hu, and W.~Zuo, ``Neural blind deconvolution using deep priors,'' in \emph{Proceedings of the IEEE/CVF Conference on Computer Vision and Pattern Recognition}, 2020, pp. 3341--3350.

\bibitem{emad2021dualsr}
M.~Emad, M.~Peemen, and H.~Corporaal, ``Dualsr: Zero-shot dual learning for real-world super-resolution,'' in \emph{Proceedings of the IEEE/CVF Winter Conference on Applications of Computer Vision}, 2021, pp. 1630--1639.

\bibitem{Yue2022blind}
Z.~Yue, Q.~Zhao, J.~Xie, L.~Zhang, D.~Meng, and K.-Y.~K. Wong, ``Blind image super-resolution with elaborate degradation modeling on noise and kernel,'' in \emph{Proceedings of the IEEE/CVF Conference on Computer Vision and Pattern Recognition}, 2022, pp. 2128--2138.

\bibitem{lai2018fast}
W.-S. Lai, J.-B. Huang, N.~Ahuja, and M.-H. Yang, ``Fast and accurate image super-resolution with deep laplacian pyramid networks,'' \emph{IEEE transactions on pattern analysis and machine intelligence}, vol.~41, no.~11, pp. 2599--2613, 2018.

\bibitem{anwar2020densely}
S.~Anwar and N.~Barnes, ``Densely residual laplacian super-resolution,'' \emph{IEEE Transactions on Pattern Analysis and Machine Intelligence}, 2020.

\bibitem{liu2020residual}
J.~Liu, W.~Zhang, Y.~Tang, J.~Tang, and G.~Wu, ``Residual feature aggregation network for image super-resolution,'' in \emph{Proceedings of the IEEE/CVF conference on computer vision and pattern recognition}, 2020, pp. 2359--2368.

\bibitem{mei2020image}
Y.~Mei, Y.~Fan, Y.~Zhou, L.~Huang, T.~S. Huang, and H.~Shi, ``Image super-resolution with cross-scale non-local attention and exhaustive self-exemplars mining,'' in \emph{Proceedings of the IEEE/CVF conference on computer vision and pattern recognition}, 2020, pp. 5690--5699.

\bibitem{huang2020unfolding}
Y.~Huang, S.~Li, L.~Wang, T.~Tan \emph{et~al.}, ``Unfolding the alternating optimization for blind super resolution,'' \emph{Advances in Neural Information Processing Systems}, vol.~33, pp. 5632--5643, 2020.

\bibitem{wang2021unsupervised}
L.~Wang, Y.~Wang, X.~Dong, Q.~Xu, J.~Yang, W.~An, and Y.~Guo, ``Unsupervised degradation representation learning for blind super-resolution,'' in \emph{Proceedings of the IEEE/CVF Conference on Computer Vision and Pattern Recognition}, 2021, pp. 10\,581--10\,590.

\bibitem{liang2021flow}
J.~Liang, K.~Zhang, S.~Gu, L.~Van~Gool, and R.~Timofte, ``Flow-based kernel prior with application to blind super-resolution,'' in \emph{Proceedings of the IEEE/CVF Conference on Computer Vision and Pattern Recognition}, 2021, pp. 10\,601--10\,610.

\bibitem{luo2022learning}
Z.~Luo, Y.~Huang, S.~Li, L.~Wang, and T.~Tan, ``Learning the degradation distribution for blind image super-resolution,'' in \emph{Proceedings of the IEEE/CVF Conference on Computer Vision and Pattern Recognition}, 2022, pp. 6063--6072.

\bibitem{kong2022reflash}
X.~Kong, X.~Liu, J.~Gu, Y.~Qiao, and C.~Dong, ``Reflash dropout in image super-resolution,'' in \emph{Proceedings of the IEEE/CVF Conference on Computer Vision and Pattern Recognition}, 2022, pp. 6002--6012.

\bibitem{saharia2022image}
C.~Saharia, J.~Ho, W.~Chan, T.~Salimans, D.~J. Fleet, and M.~Norouzi, ``Image super-resolution via iterative refinement,'' \emph{IEEE Transactions on Pattern Analysis and Machine Intelligence}, 2022.

\bibitem{fang2023self}
Z.~Fang, F.~Wu, W.~Dong, X.~Li, J.~Wu, and G.~Shi, ``Self-supervised non-uniform kernel estimation with flow-based motion prior for blind image deblurring,'' in \emph{Proceedings of the IEEE/CVF Conference on Computer Vision and Pattern Recognition}, 2023.

\bibitem{li2020continuous}
J.~Li and M.~Hu, ``Continuous model adaptation using online meta-learning for smart grid application,'' \emph{IEEE Transactions on Neural Networks and Learning Systems}, 2020.

\bibitem{chen2021multiagent}
L.~Chen, B.~Hu, Z.-H. Guan, L.~Zhao, and X.~Shen, ``Multiagent meta-reinforcement learning for adaptive multipath routing optimization,'' \emph{IEEE Transactions on Neural Networks and Learning Systems}, 2021.

\bibitem{xia2021meta}
J.~Xia and D.~Gunduz, ``Meta-learning based beamforming design for miso downlink,'' in \emph{2021 IEEE International Symposium on Information Theory (ISIT)}.\hskip 1em plus 0.5em minus 0.4em\relax IEEE, 2021, pp. 2954--2959.

\bibitem{xia2022metalearning}
J.-Y. Xia, S.~Li, J.-J. Huang, Z.~Yang, I.~M. Jaimoukha, and D.~G{\"u}nd{\"u}z, ``Metalearning-based alternating minimization algorithm for nonconvex optimization,'' \emph{IEEE Transactions on Neural Networks and Learning Systems}, 2022.

\bibitem{welling2011bayesian}
M.~Welling and Y.~W. Teh, ``Bayesian learning via stochastic gradient langevin dynamics,'' in \emph{Proceedings of the 28th international conference on machine learning (ICML-11)}, 2011, pp. 681--688.

\bibitem{arvinte2022mimo}
M.~Arvinte and J.~I. Tamir, ``Mimo channel estimation using score-based generative models,'' \emph{IEEE Transactions on Wireless Communications}, 2022.

\bibitem{russell2003exploiting}
M.~F. T. B.~C. Russell and W.~T. Freeman, ``Exploiting the sparse derivative prior for super-resolution and image demosaicing,'' in \emph{Proceedings of the Third International Workshop Statistical and Computational Theories of Vision}, 2003, pp. 1--28.

\bibitem{glasner2009super}
D.~Glasner, S.~Bagon, and M.~Irani, ``Super-resolution from a single image,'' in \emph{2009 IEEE 12th international conference on computer vision}.\hskip 1em plus 0.5em minus 0.4em\relax IEEE, 2009, pp. 349--356.

\bibitem{michaeli2013nonparametric}
T.~Michaeli and M.~Irani, ``Nonparametric blind super-resolution,'' in \emph{Proceedings of the IEEE International Conference on Computer Vision}, 2013, pp. 945--952.

\bibitem{sun2008image}
J.~Sun, Z.~Xu, and H.-Y. Shum, ``Image super-resolution using gradient profile prior,'' in \emph{2008 IEEE Conference on Computer Vision and Pattern Recognition}.\hskip 1em plus 0.5em minus 0.4em\relax IEEE, 2008, pp. 1--8.

\bibitem{krishnan2009fast}
D.~Krishnan and R.~Fergus, ``Fast image deconvolution using hyper-laplacian priors,'' \emph{Advances in neural information processing systems}, vol.~22, 2009.

\bibitem{kim2010single}
K.~I. Kim and Y.~Kwon, ``Single-image super-resolution using sparse regression and natural image prior,'' \emph{IEEE transactions on pattern analysis and machine intelligence}, vol.~32, no.~6, pp. 1127--1133, 2010.

\bibitem{rudin1992nonlinear}
L.~I. Rudin, S.~Osher, and E.~Fatemi, ``Nonlinear total variation based noise removal algorithms,'' \emph{Physica D: nonlinear phenomena}, vol.~60, no. 1-4, pp. 259--268, 1992.

\bibitem{jin2018normalized}
M.~Jin, S.~Roth, and P.~Favaro, ``Normalized blind deconvolution,'' in \emph{Proceedings of the European Conference on Computer Vision (ECCV)}, 2018, pp. 668--684.

\bibitem{dong2014learning}
C.~Dong, C.~C. Loy, K.~He, and X.~Tang, ``Learning a deep convolutional network for image super-resolution,'' in \emph{European conference on computer vision}.\hskip 1em plus 0.5em minus 0.4em\relax Springer, 2014, pp. 184--199.

\bibitem{kim2016accurate}
J.~Kim, J.~K. Lee, and K.~M. Lee, ``Accurate image super-resolution using very deep convolutional networks,'' in \emph{Proceedings of the IEEE conference on computer vision and pattern recognition}, 2016, pp. 1646--1654.

\bibitem{zhang2018learning}
K.~Zhang, W.~Zuo, and L.~Zhang, ``Learning a single convolutional super-resolution network for multiple degradations,'' in \emph{Proceedings of the IEEE Conference on Computer Vision and Pattern Recognition}, 2018, pp. 3262--3271.

\bibitem{xu2020unified}
Y.-S. Xu, S.-Y.~R. Tseng, Y.~Tseng, H.-K. Kuo, and Y.-M. Tsai, ``Unified dynamic convolutional network for super-resolution with variational degradations,'' in \emph{Proceedings of the IEEE/CVF Conference on Computer Vision and Pattern Recognition}, 2020, pp. 12\,496--12\,505.

\bibitem{hui2021learning}
Z.~Hui, J.~Li, X.~Wang, and X.~Gao, ``Learning the non-differentiable optimization for blind super-resolution,'' in \emph{Proceedings of the IEEE/CVF conference on computer vision and pattern recognition}, 2021, pp. 2093--2102.

\bibitem{kim2021koalanet}
S.~Y. Kim, H.~Sim, and M.~Kim, ``Koalanet: Blind super-resolution using kernel-oriented adaptive local adjustment,'' in \emph{Proceedings of the IEEE/CVF conference on computer vision and pattern recognition}, 2021, pp. 10\,611--10\,620.

\bibitem{zhang2021designing}
K.~Zhang, J.~Liang, L.~Van~Gool, and R.~Timofte, ``Designing a practical degradation model for deep blind image super-resolution,'' in \emph{Proceedings of the IEEE/CVF International Conference on Computer Vision}, 2021, pp. 4791--4800.

\bibitem{fang2022uncertainty}
Z.~Fang, W.~Dong, X.~Li, J.~Wu, L.~Li, and G.~Shi, ``Uncertainty learning in kernel estimation for multi-stage blind image super-resolution,'' in \emph{European Conference on Computer Vision}.\hskip 1em plus 0.5em minus 0.4em\relax Springer, 2022, pp. 144--161.

\bibitem{xia2022knowledge}
B.~Xia, Y.~Zhang, Y.~Wang, Y.~Tian, W.~Yang, R.~Timofte, and L.~Van~Gool, ``Knowledge distillation based degradation estimation for blind super-resolution,'' \emph{arXiv preprint arXiv:2211.16928}, 2022.

\bibitem{zheng2022unfolded}
H.~Zheng, H.~Yong, and L.~Zhang, ``Unfolded deep kernel estimation for blind image super-resolution,'' in \emph{European Conference on Computer Vision}.\hskip 1em plus 0.5em minus 0.4em\relax Springer, 2022, pp. 502--518.

\bibitem{liang2021mutual}
J.~Liang, G.~Sun, K.~Zhang, L.~Van~Gool, and R.~Timofte, ``Mutual affine network for spatially variant kernel estimation in blind image super-resolution,'' in \emph{Proceedings of the IEEE/CVF International Conference on Computer Vision}, 2021, pp. 4096--4105.

\bibitem{hu2019meta}
X.~Hu, H.~Mu, X.~Zhang, Z.~Wang, T.~Tan, and J.~Sun, ``Meta-sr: A magnification-arbitrary network for super-resolution,'' in \emph{Proceedings of the IEEE/CVF conference on computer vision and pattern recognition}, 2019, pp. 1575--1584.

\bibitem{soh2020meta}
J.~W. Soh, S.~Cho, and N.~I. Cho, ``Meta-transfer learning for zero-shot super-resolution,'' in \emph{Proceedings of the IEEE/CVF Conference on Computer Vision and Pattern Recognition}, 2020, pp. 3516--3525.

\bibitem{li2021different}
H.~Li, Y.~Cen, Y.~Liu, X.~Chen, and Z.~Yu, ``Different input resolutions and arbitrary output resolution: a meta learning-based deep framework for infrared and visible image fusion,'' \emph{IEEE Transactions on Image Processing}, vol.~30, pp. 4070--4083, 2021.

\bibitem{yu2021lite}
C.~Yu, B.~Xiao, C.~Gao, L.~Yuan, L.~Zhang, N.~Sang, and J.~Wang, ``Lite-hrnet: A lightweight high-resolution network,'' in \emph{Proceedings of the IEEE/CVF Conference on Computer Vision and Pattern Recognition}, 2021, pp. 10\,440--10\,450.

\bibitem{xia2022metaBSR}
B.~Xia, Y.~Tian, Y.~Zhang, Y.~Hang, W.~Yang, and Q.~Liao, ``Meta-learning based degradation representation for blind super-resolution,'' \emph{arXiv preprint arXiv:2207.13963}, 2022.

\bibitem{cornillere2019blind}
V.~Cornillere, A.~Djelouah, W.~Yifan, O.~Sorkine-Hornung, and C.~Schroers, ``Blind image super-resolution with spatially variant degradations,'' \emph{ACM Transactions on Graphics (TOG)}, vol.~38, no.~6, pp. 1--13, 2019.

\bibitem{zhang2019deep}
K.~Zhang, W.~Zuo, and L.~Zhang, ``Deep plug-and-play super-resolution for arbitrary blur kernels,'' in \emph{Proceedings of the IEEE/CVF Conference on Computer Vision and Pattern Recognition}, 2019, pp. 1671--1681.

\bibitem{zhang2020deep}
K.~Zhang, L.~V. Gool, and R.~Timofte, ``Deep unfolding network for image super-resolution,'' in \emph{Proceedings of the IEEE/CVF conference on computer vision and pattern recognition}, 2020, pp. 3217--3226.

\bibitem{li2020efficient}
Y.~Li, M.~Tofighi, J.~Geng, V.~Monga, and Y.~C. Eldar, ``Efficient and interpretable deep blind image deblurring via algorithm unrolling,'' \emph{IEEE Transactions on Computational Imaging}, vol.~6, pp. 666--681, 2020.

\bibitem{fu2022kxnet}
J.~Fu, H.~Wang, Q.~Xie, Q.~Zhao, D.~Meng, and Z.~Xu, ``Kxnet: A model-driven deep neural network for blind super-resolution,'' in \emph{European Conference on Computer Vision}.\hskip 1em plus 0.5em minus 0.4em\relax Springer, 2022, pp. 235--253.

\bibitem{gu2019blind}
J.~Gu, H.~Lu, W.~Zuo, and C.~Dong, ``Blind super-resolution with iterative kernel correction,'' in \emph{Proceedings of the IEEE/CVF Conference on Computer Vision and Pattern Recognition}, 2019, pp. 1604--1613.

\bibitem{li2020boosting}
A.~Li, W.~Huang, X.~Lan, J.~Feng, Z.~Li, and L.~Wang, ``Boosting few-shot learning with adaptive margin loss,'' in \emph{Proceedings of the IEEE/CVF conference on computer vision and pattern recognition}, 2020, pp. 12\,576--12\,584.

\bibitem{Yang2022ALearning}
Z.~Yang, J.-Y. Xia, J.~Luo, S.~Zhang, and D.~G\"und\"uz, ``A learning aided flexible gradient descent approach to miso beamforming,'' \emph{IEEE Wireless Communications Letters}, pp. 1--1, 2022.

\bibitem{elad1997restoration}
M.~Elad and A.~Feuer, ``Restoration of a single superresolution image from several blurred, noisy, and undersampled measured images,'' \emph{IEEE transactions on image processing}, vol.~6, no.~12, pp. 1646--1658, 1997.

\bibitem{farsiu2004advances}
S.~Farsiu, D.~Robinson, M.~Elad, and P.~Milanfar, ``Advances and challenges in super-resolution,'' \emph{International Journal of Imaging Systems and Technology}, vol.~14, no.~2, pp. 47--57, 2004.

\bibitem{liu2013bayesian}
C.~Liu and D.~Sun, ``On bayesian adaptive video super resolution,'' \emph{IEEE transactions on pattern analysis and machine intelligence}, vol.~36, no.~2, pp. 346--360, 2013.

\bibitem{zhang2018image}
Y.~Zhang, K.~Li, K.~Li, L.~Wang, B.~Zhong, and Y.~Fu, ``Image super-resolution using very deep residual channel attention networks,'' in \emph{Proceedings of the European conference on computer vision (ECCV)}, 2018, pp. 286--301.

\bibitem{gandelsman2019double}
Y.~Gandelsman, A.~Shocher, and M.~Irani, ``" double-dip": Unsupervised image decomposition via coupled deep-image-priors,'' in \emph{Proceedings of the IEEE/CVF Conference on Computer Vision and Pattern Recognition}, 2019, pp. 11\,026--11\,035.

\bibitem{kingma2014adam}
D.~P. Kingma and J.~Ba, ``Adam: A method for stochastic optimization,'' \emph{arXiv preprint arXiv:1412.6980}, 2014.

\bibitem{riegler2015conditioned}
G.~Riegler, S.~Schulter, M.~Ruther, and H.~Bischof, ``Conditioned regression models for non-blind single image super-resolution,'' in \emph{Proceedings of the IEEE International Conference on Computer Vision}, 2015, pp. 522--530.

\bibitem{shao2015simple}
W.-Z. Shao and M.~Elad, ``Simple, accurate, and robust nonparametric blind super-resolution,'' in \emph{Image and Graphics}.\hskip 1em plus 0.5em minus 0.4em\relax Springer, 2015, pp. 333--348.

\bibitem{shocher2018zero}
A.~Shocher, N.~Cohen, and M.~Irani, ``“zero-shot” super-resolution using deep internal learning,'' in \emph{Proceedings of the IEEE conference on computer vision and pattern recognition}, 2018, pp. 3118--3126.

\bibitem{kupyn2018deblurgan}
O.~Kupyn, V.~Budzan, M.~Mykhailych, D.~Mishkin, and J.~Matas, ``Deblurgan: Blind motion deblurring using conditional adversarial networks,'' in \emph{Proceedings of the IEEE conference on computer vision and pattern recognition}, 2018, pp. 8183--8192.

\bibitem{bevilacqua2012low}
M.~Bevilacqua, A.~Roumy, C.~Guillemot, and M.~L. Alberi-Morel, ``Low-complexity single-image super-resolution based on nonnegative neighbor embedding,'' in \emph{British Machine Vision Conference}, 2012, pp. 135--1.

\bibitem{zeyde2010single}
R.~Zeyde, M.~Elad, and M.~Protter, ``On single image scale-up using sparse-representations,'' in \emph{International conference on curves and surfaces}.\hskip 1em plus 0.5em minus 0.4em\relax Springer, 2010, pp. 711--730.

\bibitem{martin2001database}
D.~Martin, C.~Fowlkes, D.~Tal, and J.~Malik, ``A database of human segmented natural images and its application to evaluating segmentation algorithms and measuring ecological statistics,'' in \emph{Proceedings Eighth IEEE International Conference on Computer Vision. ICCV 2001}, vol.~2.\hskip 1em plus 0.5em minus 0.4em\relax IEEE, 2001, pp. 416--423.

\bibitem{huang2015single}
J.-B. Huang, A.~Singh, and N.~Ahuja, ``Single image super-resolution from transformed self-exemplars,'' in \emph{Proceedings of the IEEE conference on computer vision and pattern recognition}, 2015, pp. 5197--5206.

\bibitem{wang2004image}
Z.~Wang, A.~C. Bovik, H.~R. Sheikh, and E.~P. Simoncelli, ``Image quality assessment: from error visibility to structural similarity,'' \emph{IEEE transactions on image processing}, vol.~13, no.~4, pp. 600--612, 2004.

\end{thebibliography}

\begin{IEEEbiography}
[{\includegraphics[width=0.5in,height=0.7in,keepaspectratio]{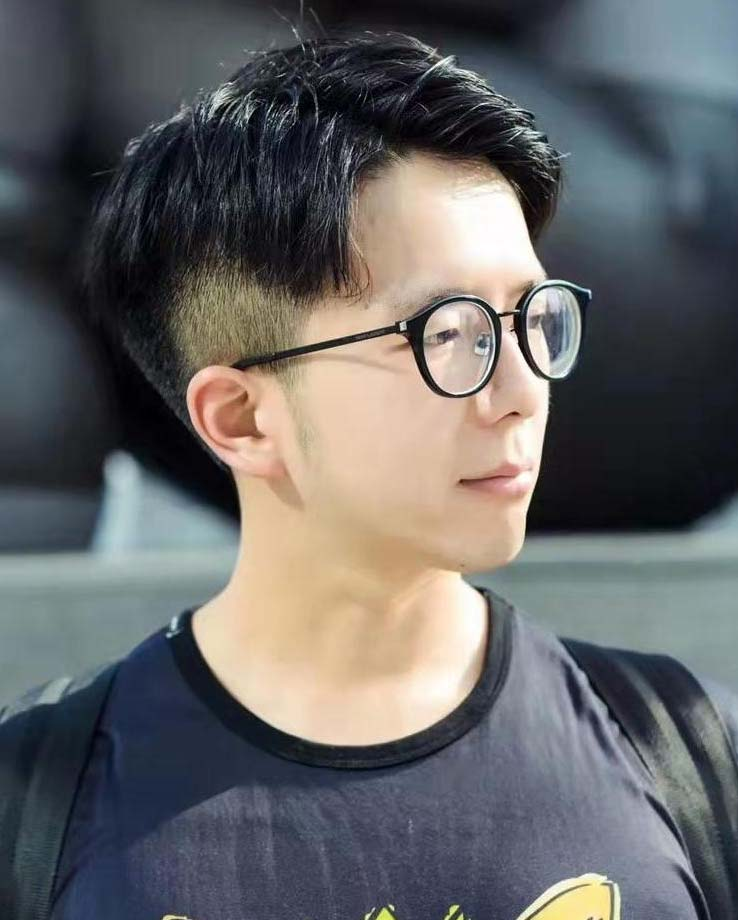}}]
{Jingyuan Xia} is currently an Associate Professor with the College of the Electronic Science, the National University of Defense Technology (NUDT). His current research interests include low level image processing, nonconvex optimization, and machine learning for signal processing.
\end{IEEEbiography}

\begin{IEEEbiography}[{\includegraphics[width=0.5in,height=0.7in,keepaspectratio]{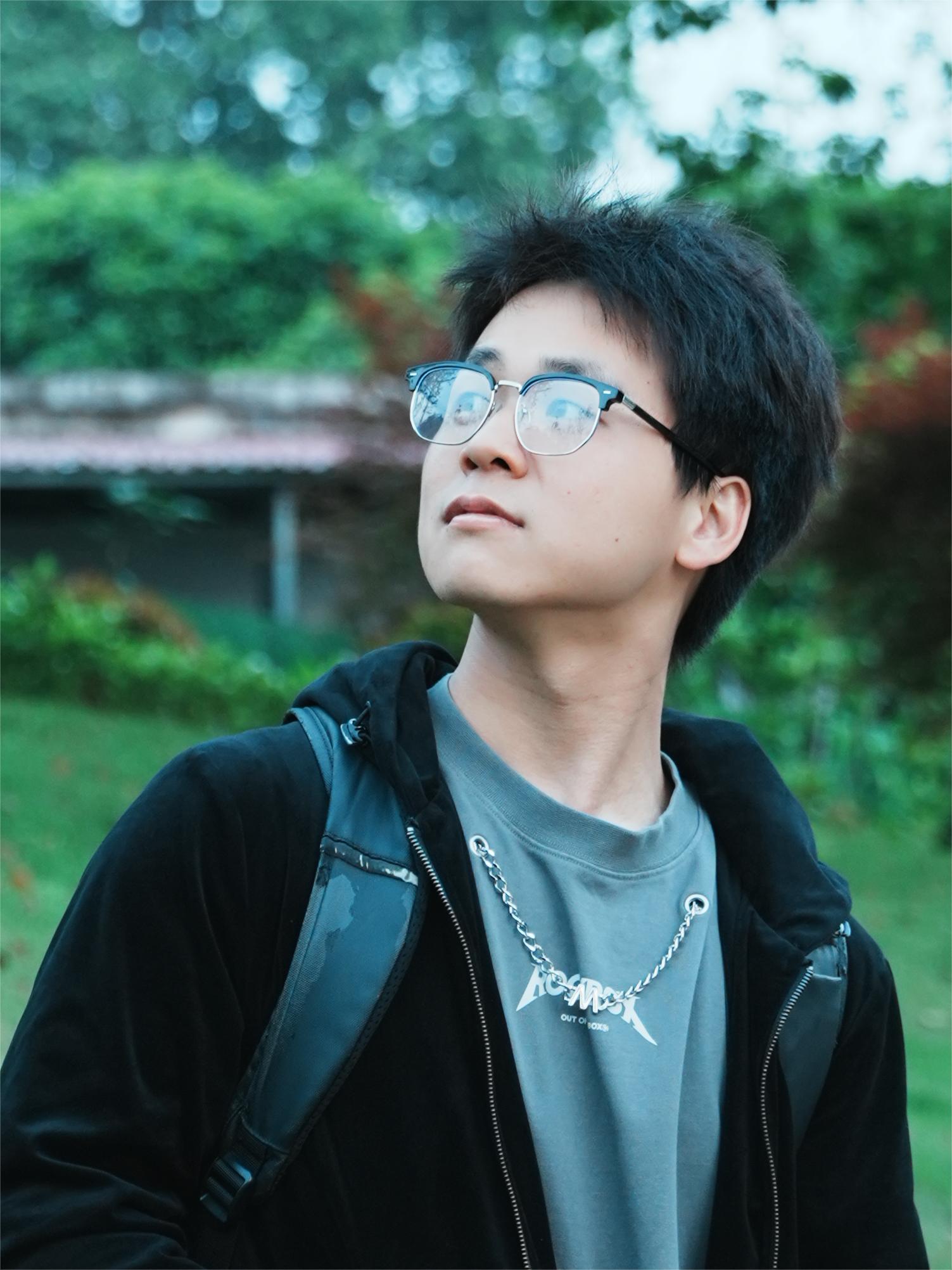}}]
{Zhixiong Yang} 
is now pursuing a Ph.D. degree at the College of Electronic Science, NUDT. 
His research interests include low-level image restoration and ISAR imaging.
\end{IEEEbiography}

\begin{IEEEbiography}[{\includegraphics[width=0.5in,height=0.7in,keepaspectratio]{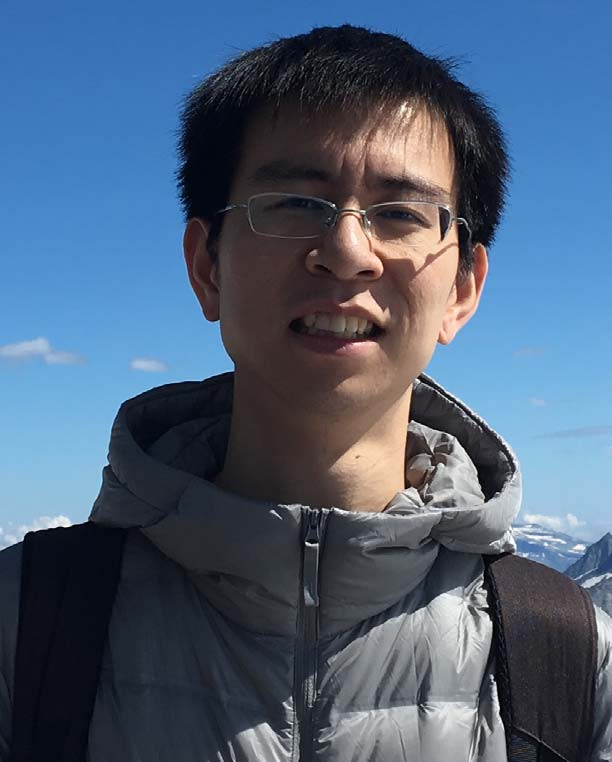}}]
{Shengxi Li} (Member, IEEE) is a Professor with Beihang University. His research interests include generative models, statistical signal processing, rate-distortion theory, and perceptual video coding.
\end{IEEEbiography}

\begin{IEEEbiography}[{\includegraphics[width=0.5in,height=0.7in,keepaspectratio]{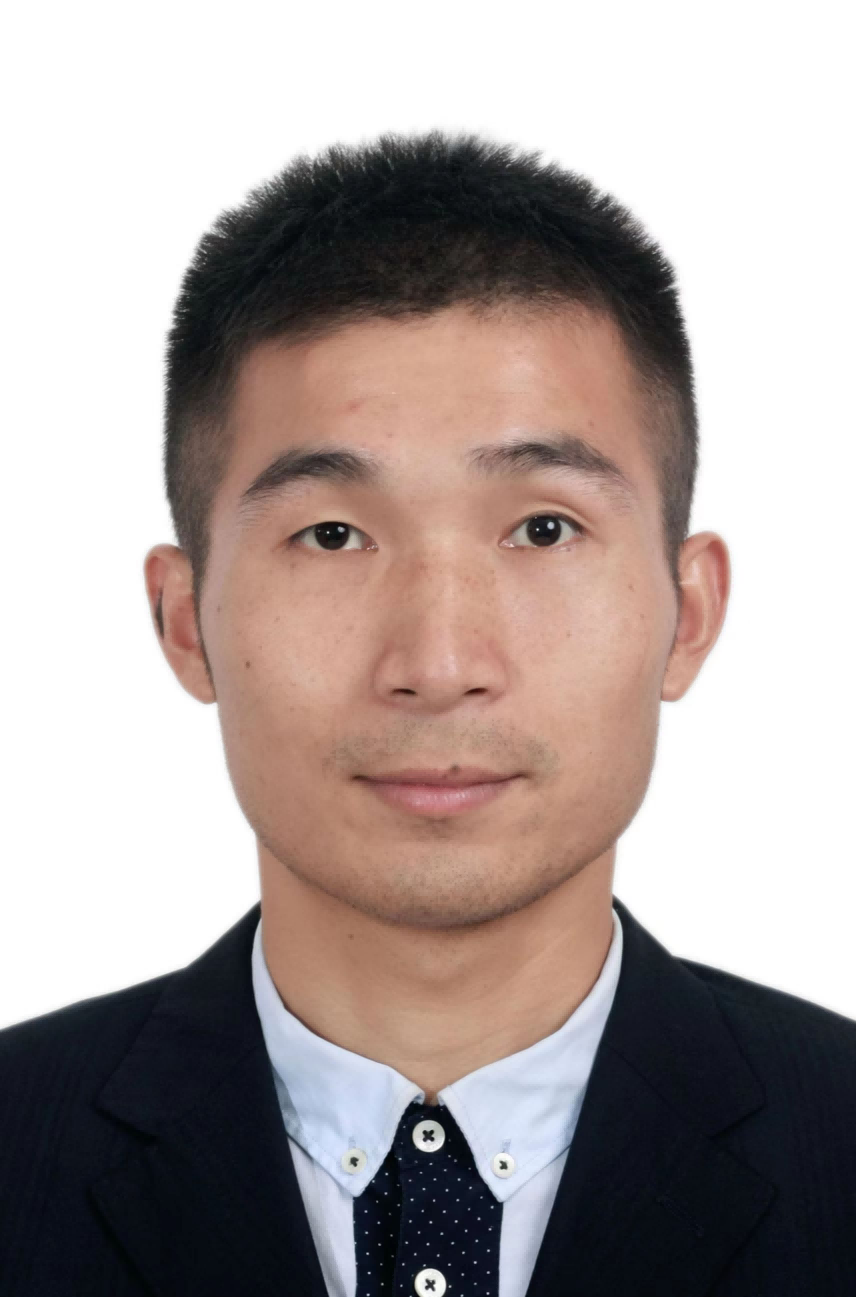}}]
{Shuanghui Zhang} (Member, IEEE) is currently a Professor with the College of Electrical Science and Technology, NUDT. His research interests lie in compressive sensing, sparse signal recovery techniques, Bayesian inference, and their applications in radar signal processing.
\end{IEEEbiography}


\begin{IEEEbiography}[{\includegraphics[width=0.5in,height=0.7in,keepaspectratio]{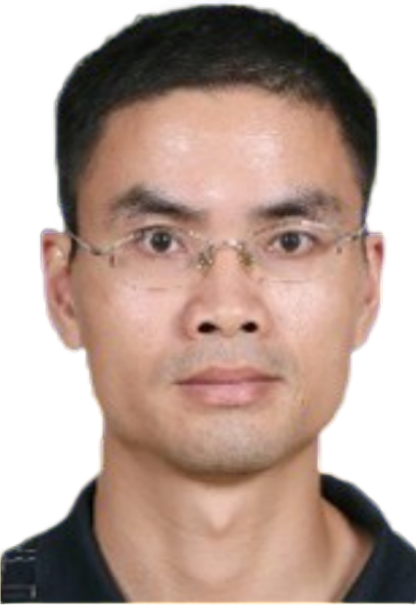}}]
{Yaowen Fu} is currently a Professor with the College of Electronic Science and Technology, NUDT. His research interests include information fusion, radar signal processing, and other aspects of research.
\end{IEEEbiography}

\begin{IEEEbiography}[{\includegraphics[width=0.5in,height=0.7in,keepaspectratio]{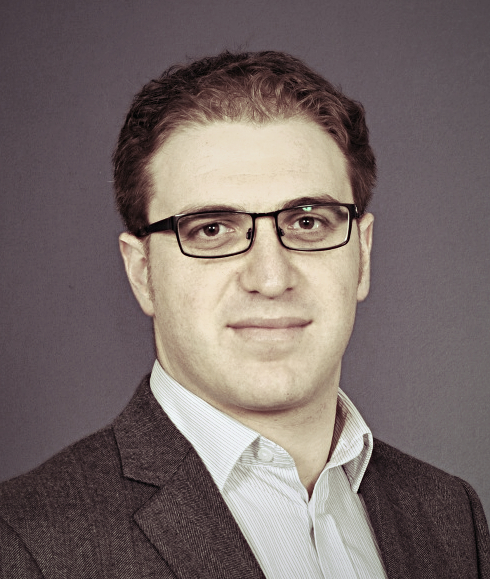}}]
{Deniz Gündüz} 
is a Fellow of the IEEE, and a Distinguished Lecturer for the IEEE Information Theory Society (2020-22). He is the recipient of a Consolidator Grant of the European Research Council (ERC) in 2022, the IEEE Communications Society - Communication Theory Technical Committee (CTTC) Early Achievement Award in 2017, Starting and Consolidator Grants of the European Research Council (ERC), and several best paper awards. He is currently a Professor in Imperial College London.
\end{IEEEbiography}



\begin{IEEEbiography}[{\includegraphics[width=0.5in,height=0.7in,keepaspectratio]{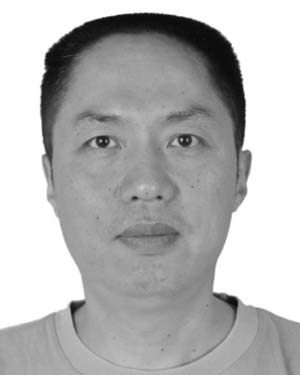}}]
{Xiang Li} was elected as the Academician of the Chinese Academy of Sciences in 2022, and a Professor with NUDT. His research interests include signal processing, automation target recognition, and machine learning.
\end{IEEEbiography}

\end{document}